\documentclass[PAPER, atlasdraft=false, USenglish, cernpreprint, texmf, orcidlogo]{atlasdoc}
\usepackage{atlaspackage}
\usepackage{comment}
\usepackage{multirow}
\usepackage{atlasbiblatex}

\usepackage[jetetmiss]{atlasphysics}
\usepackage{cleveref}
\usepackage{pgffor} %

\graphicspath{{logos/}{figures/}}

\usepackage{ANA-HIGP-2024-31-PAPER-defs}

\AtlasTitle{Search for new scalars via \(X \rightarrow SH \rightarrow b\bar{b}b\bar{b}\) in proton--proton collisions at \(\sqrt{s} = 13\)~TeV with the ATLAS detector}

\AtlasVersion{2.1}

\AtlasAbstract{%

A search is performed for a scalar resonance $X$ decaying into a Standard Model Higgs boson $H$ and an additional scalar $S$, each decaying into a bottom--antibottom-quark pair, giving a $b\bar{b}b\bar{b}$ final state. Two event topologies are considered, depending on whether the Higgs boson is reconstructed from a single large-radius jet or from a pair of small-radius jets. The analysis uses proton--proton collision data at $\sqrt{s} = 13$~\TeV, collected by the ATLAS experiment at the Large Hadron Collider during \RunTwo, corresponding to an integrated luminosity of $126$--$136~\ifb$ depending on the topology and the associated triggers. The search covers masses of 300~\GeV--6~\TeV\ for $X$ and 70~\GeV--5~\TeV\ for $S$.  The dominant multijet and $t\bar{t}$ backgrounds are estimated with a hierarchical model using Gaussian processes and normalizing flows. No significant excess over the Standard Model background expectation is observed. Upper limits at 95\% confidence level are set on the production cross-section times branching ratio, ranging from $0.7$~fb to $2.6$~pb.
}

\AtlasRefCode{HIGP-2024-31}

\PreprintIdNumber{CERN-EP-2026-202}

\AtlasJournal{Phys.\ Rev.\ D.}
\AtlasCoverEgroupAnalysisTeam{atlas-higp-2024-31-analysis-team@cern.ch}

\hypersetup{pdftitle={ATLAS document},pdfauthor={The ATLAS Collaboration}}

\begin{document}

\maketitle

\tableofcontents

\newpage

% The next lines are included from the .//sections/intro.tex input file
\section{Introduction}
\label{sec:intro}

The discovery of the Higgs boson in 2012 by the ATLAS and CMS experiments at the Large Hadron Collider (LHC) marked a milestone in particle physics and opened the door to exploring its properties~\cite{HIGG-2012-27,CMS-HIG-12-028}. Subsequent measurements have shown no significant deviations from Standard Model (SM) expectations~\cite{HIGG-2021-23,CMS-HIG-22-001}, but current experimental constraints still leave room for effects from physics beyond the SM (BSM). Despite its remarkable success, the SM does not account for several observed phenomena, such as dark matter and neutrino masses, and does not provide a satisfactory explanation of naturalness, baryogenesis, or gravity.

Extensions of the Higgs sector, with sizable viable parameter space, provide a well-motivated framework to probe BSM physics at the LHC\@. Such extensions range from minimal scalar-singlet models~\cite{bosse2026}, through two-Higgs-doublet~\cite{Branco_2012} and inert-doublet scenarios~\cite{Belyaev_2018}, to singlet-extended two-doublet frameworks~\cite{Arhrib_2020}, and more elaborate multi-doublet~\cite{Darvishi_2021} or triplet Higgs sectors~\cite{GM}. These models predict the existence of additional scalar states that can couple to the SM Higgs boson and give rise to distinctive experimental signatures. In this context, the Higgs boson becomes a powerful tool in searches for new physics. Searches for new resonances decaying into Higgs boson pairs have been performed extensively in multiple final states, in particular in the $b\bar{b}b\bar{b}$ final state, denoted $4b$ in the remainder of this paper, which provides the largest branching ratio, by ATLAS~\cite{HDBS-2018-41,HDBS-2022-02} and CMS~\cite{CMS-B2G-20-004,CMS-B2G-21-003,CMS-HIG-24-010}, with no evidence for new physics observed to date.

The asymmetric decay of a heavy scalar resonance $X$ into a SM Higgs boson $H$ and a new scalar $S$ provides a complementary and well-motivated search avenue for extended Higgs sectors. Depending on the model parameters, the decay $X\to SH$ can yield larger signal rates than the symmetric decay $X\to HH$. Such signatures are predicted by many BSM scenarios, including the Two-Real-Singlet Model (TRSM), which expands the SM with two additional scalar fields~\cite{Robens_2020}. After mixing with the SM Higgs doublet, three scalars with a mass hierarchy emerge, one of which can be identified with the observed Higgs boson, while the other two correspond to the $X$ and $S$ states. The free parameters of the model include the vacuum expectation values associated with the new singlets ($v_X$ and $v_S$), and the mixing angles between them ($\theta_{HS}$, $\theta_{HX}$, $\theta_{XS}$). Similar signatures are also predicted in other models, such as the CP-violating Two-Higgs-Doublet Model (2HDM)~\cite{Branco_2012} and the next-to-minimal supersymmetric extension of the SM (NMSSM)~\cite{Ellwanger_2010}, which introduce an additional doublet and a singlet field.

The $X \to SH$ process has been probed at the LHC in several final states. The CMS Collaboration has published several searches for the process in final states including $b\bar{b}b\bar{b}$~\cite{CMS-B2G-21-003, CMS-HIG-20-012}, $bb4q$~\cite{CMS-B2G-23-007}, $bb\gamma\gamma$~\cite{CMS-B2G-24-001}, and $bbZZ$~\cite{CMS-B2G-24-022}, and the ATLAS Collaboration has searched in the $SH \to HHH \to 6b$~\cite{HIGP-2024-32}, $VV\tau\tau$~\cite{HDBS-2022-44}, $bb\gamma\gamma$~\cite{HIGP-2024-34_copy}, and $\gamma\gamma VV$~\cite{HDBS-2021-23} final states. The CMS $X \to SH$ search in the resolved $4b$ final state reported its largest deviation from the background-only hypothesis at $(m_{\mathrm{X}}, m_{\mathrm{S}}) = (\text{600\,GeV}, \text{400\,GeV})$, with a local (global) significance of 3.47 (2.44) standard deviations~\cite{CMS-HIG-20-012}.

\begin{figure}[t]
\center
\includegraphics[width=.35\textwidth]{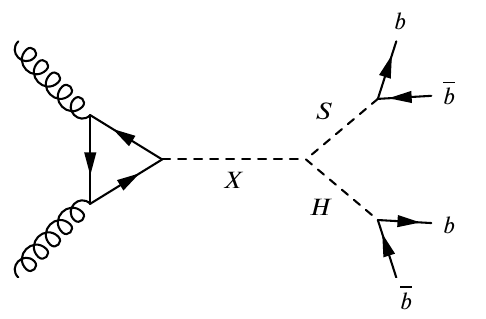}
\caption{Leading-order Feynman diagram of the signal production and decay process. The heavy scalar $X$ is produced via gluon--gluon fusion and decays via $X\to SH\to 4b$, where $S$ is a scalar lighter than $X$ and $H$ is the SM-like Higgs boson.}
\label{fig:feynman}
\end{figure}

This paper reports the first ATLAS search for $X\to SH\to 4b$, performed using the \RunTwo data, where the heavy spin-0 resonance $X$ is produced via gluon--gluon fusion, as illustrated in \cref{fig:feynman}. The search probes mass ranges of $300~\GeV$--$6~\TeV$ for $X$ and $70~\GeV$--$5~\TeV$ for $S$. The wide mass range leads to two distinct kinematic regimes, depending on the mass difference between $X$ and $H$. Accordingly, two complementary topologies are identified. In the \emph{resolved} topology, corresponding to small or moderate mass differences between $X$ and $S$, both the $H$ and $S$ bosons are reconstructed from pairs of \bjets. In the \emph{mixed} topology (boosted $H$, resolved $S$), relevant when $X$ is significantly heavier than $S$, the Higgs boson is produced with large transverse momentum \pT, and its decay products are reconstructed as a single large-radius (\emph{boosted}) jet, while the $S$ boson is reconstructed from a pair of \bjets, as in the resolved case. Two analysis channels are defined to target these distinct topologies to maximize the overall sensitivity. The fully boosted topology, where both $H$ and $S$ are reconstructed as large-radius jets, is not considered in this work but is left for a future dedicated search.

The paper is organized as follows. A brief overview of the ATLAS detector is given in \cref{sec:detector}, and a description of the data and simulated samples used in the analysis is provided in \cref{sec:samples}. The physics objects, including the advanced transformer-based algorithms used to identify b-jets in the resolved and boosted regimes, GN2 and GN2X~\cite{vaswani2023attentionneed, FTAG-2023-05, ATL-PHYS-PUB-2023-021}, are described in \cref{sec:objects}. The event selection is detailed in \cref{sec:selection}. The fully data-driven background estimate, which models the dominant multijet and $t\bar{t}$ backgrounds in signal-depleted control regions using Gaussian processes~\cite{Rasmussen} and normalizing flows~\cite{Daumann_2024}, is summarized in \cref{sec:background}. The assessment of systematic uncertainties is discussed in \cref{sec:systematics}. \cref{sec:results} presents the statistical analysis, based on a profile likelihood fit to the reconstructed $X$ mass distribution, and the results, interpreted as 95\% CL upper limits on $\sigma \times \mathcal{B}\left(X\to SH \to 4b\right)$ and compared with TRSM predictions. The conclusions are given in \cref{sec:conclusion}.

% End of text imported from the .//sections/intro.tex input file

% The next lines are included from the .//sections/detector.tex input file
\section{ATLAS detector}
\label{sec:detector}

\newcommand{\AtlasCoordFootnote}{%
ATLAS uses a right-handed coordinate system with its origin at the nominal interaction point (IP) in the center of the detector and the \(z\)-axis along the beam pipe. The \(x\)-axis points from the IP to the center of the LHC ring, and the \(y\)-axis points upwards. Polar coordinates \((r,\phi)\) are used in the transverse plane, \(\phi\) being the azimuthal angle around the \(z\)-axis. The pseudorapidity is defined in terms of the polar angle \(\theta\) as \(\eta = -\ln \tan(\theta/2)\) and is equal to the rapidity $ y = \frac{1}{2} \ln \left( \frac{E + p_z}{E - p_z} \right) $ in the relativistic limit. Angular distance is measured in units of \(\Delta R \equiv \sqrt{(\Delta y)^{2} + (\Delta\phi)^{2}}\).}

The ATLAS detector~\cite{PERF-2007-01} at the LHC covers nearly the entire solid angle around the collision point.\footnote{\AtlasCoordFootnote}
It consists of an inner tracking detector surrounded by a thin superconducting solenoid, electromagnetic and hadronic calorimeters,
and a muon spectrometer incorporating three large superconducting air-core toroidal magnets.

The inner-detector system (ID) is immersed in a \qty{2}{\tesla} axial magnetic field
and provides charged-particle tracking in the range of \(|\eta| < 2.5\).
The high-granularity silicon pixel detector covers the vertex region and typically provides four measurements per track,
the first hit generally being in the insertable B-layer (IBL) installed before Run~2~\cite{ATLAS-TDR-19,PIX-2018-001}.
It is followed by the Semi-Conductor Tracker (SCT), which usually provides eight measurements per track.
These silicon detectors are complemented by the transition radiation tracker (TRT),
which enables radially extended track reconstruction up to \(|\eta| = 2.0\).
The TRT also provides electron identification information
based on the fraction of hits (typically 30 in total) above a higher energy-deposit threshold corresponding to transition radiation.

The calorimeter system covers the pseudorapidity range \(|\eta| < 4.9\).
Within the region \(|\eta|< 3.2\), electromagnetic calorimetry is provided by barrel and
endcap high-granularity lead/liquid-argon (LAr) calorimeters,
with an additional thin LAr presampler covering \(|\eta| < 1.8\)
to correct for energy loss in material upstream of the calorimeters.
Hadronic calorimetry is provided by the steel/scintillator-tile calorimeter,
segmented into three barrel structures within \(|\eta| < 1.7\), and two copper/LAr hadronic endcap calorimeters.
The solid angle coverage is completed with forward copper/LAr and tungsten/LAr calorimeter modules
optimized for electromagnetic and hadronic energy measurements, respectively.

The muon spectrometer (MS) comprises separate trigger and
high-precision tracking chambers measuring the deflection of muons in a magnetic field generated by the superconducting air-core toroidal magnets.
The field integral of the toroids ranges between \num{2.0} and \qty{6.0}{\tesla\metre}
across most of the detector.
Three layers of precision chambers, each consisting of layers of monitored drift tubes, cover the region \(|\eta| < 2.7\),
complemented by cathode-strip chambers in the forward region, where the background is highest.
The muon trigger system covers the range \(|\eta| < 2.4\) with resistive-plate chambers in the barrel, and thin-gap chambers in the endcap regions.

The luminosity is measured mainly by the LUCID--2~\cite{LUCID2} detector that records Cherenkov light produced in the quartz windows of photomultipliers located close to the beampipe.

Events were selected by the first-level trigger system implemented in custom hardware,
followed by selections made by algorithms implemented in software in the high-level trigger~\cite{TRIG-2016-01}.
The first-level trigger accepted events from the \qty{40}{\MHz} bunch crossings at a rate close to \qty{100}{\kHz},
which the high-level trigger further reduced to record complete events to disk at about \qty{1.25}{\kHz}.

A software suite~\cite{SOFT-2022-02} is used in data simulation, in the reconstruction
and analysis of real and simulated data, in detector operations, and in the trigger and data acquisition
systems of the experiment.

% End of text imported from the .//sections/detector.tex input file

% The next lines are included from the .//sections/samples.tex input file
\section{Data and simulated event samples}
\label{sec:samples}

\subsection{Data and triggers}

This search uses $pp$ collision data collected between 2016 and 2018 during \RunTwo at $\sqrt{s}=13~\TeV$, corresponding to an integrated luminosity of $126$--$136~\ifb$~\cite{DAPR-2021-01}. The luminosity difference arises from the channel-specific triggers. Only data collected under stable beam conditions, with all relevant detector systems functional, are used~\cite{DAPR-2018-01}.

Events are selected using triggers requiring online \btag of jets reconstructed with the anti-$k_t$ algorithm ~\cite{Cacciari:2008gp} with a radius parameter $R=0.4$~\cite{Fastjet} (\bjet triggers)~\cite{TRIG-2018-08}.
The mixed channel also includes jet triggers requiring the presence of a large-radius (\largeR) jet with $R=1.0$.
Both channels adopt a \emph{trigger bucket} approach to simplify the application of the trigger scale factors (SFs). Events are assigned exclusively to a single bucket using offline selection criteria.
The buckets are chosen to maximize trigger efficiency independently for resolved and mixed channels.

The resolved channel uses two trigger buckets, denoted \texttt{2b+1j} and \texttt{2b+2j}. Events containing at least three jets with $\pt > 70$~\GeV, one of which also satisfies $\pt > 170$~\GeV, are assigned to the \texttt{2b+1j} bucket if they pass the corresponding trigger. Remaining events are assigned to the \texttt{2b+2j} bucket if they pass its corresponding trigger.

The online \btag relies on the multivariate algorithm \emph{MV2}~\cite{TRIG-2018-08}, which uses a boosted decision tree (BDT). Several $b$-jet triggers are used, with selection criteria and \btag\ working points that depend on the data-taking year, retaining approximately 40--70\% of true $b$-jets in the corresponding reference samples.

In the \texttt{2b+1j} bucket, the trigger requires two $b$-tagged jets with $\ET>55~\GeV$ and one additional jet with $\ET>110~\GeV$ in 2016 and $\ET>150~\GeV$ in 2017--2018 seeded with L1 jets within $|\eta| < 3.2$. The \btag\ algorithms consider jets with $|\eta| < 2.5$ and operate at a working point of 60\% in 2016 and 70\% in 2017--2018.
In the \texttt{2b+2j} bucket, the trigger requires two $b$-tagged jets with $\ET>35~\GeV$, together with two additional jets satisfying the same $\ET$ requirement, seeded with L1 jets within $\eta < 2.5$. The \btag\ working point is 60\%, tightened to 40\% in 2017.

The mixed channel also uses two trigger buckets, denoted \texttt{1Lj} and \texttt{2b+1j}. If the event contains a large-$R$ jet with $\pt>520~\GeV$, jet mass greater than $70~\GeV$, and $|\eta|<2.0$, it is assigned to the \texttt{1Lj} bucket. In this case, the event is required to satisfy a jet trigger based on a large-radius jet with $\ET>420~\GeV$ at the high-level trigger (HLT), seeded by a $100~\GeV$ jet at L1. The offline threshold is set above the online one to ensure operation on the trigger efficiency plateau. If no such large-$R$ jet is present, the event is considered in the \texttt{2b+1j} bucket, which requires the same triggers as the \texttt{2b+1j} bucket of the resolved channel.

For the \texttt{2b+1j} and \texttt{2b+2j} buckets in both channels,  the offline jet selections are such that the analysis operates in the region where the trigger efficiency rises with the jet $E_{\mathrm{T}}$. Dedicated trigger SFs measured in di-leptonic $t\bar{t}$ events are applied to take into account different trigger efficiencies between data and simulation.

Typical trigger efficiencies for the signal samples in the resolved channel are around 90\%, and they reach almost 100\% for the mixed channel at high $X$ masses. These efficiencies are calculated relative to the offline event selection (Section~\ref{sec:selection}).

The 2016 data are affected by inefficiency in the online primary-vertex reconstruction, which reduces the efficiency of the $b$-jet trigger selection~\cite{TRIG-2018-08}. Additionally, in all years, luminosity blocks at the start of each run are discarded due to beamspot calibration requirements. As a result, the integrated luminosity associated with events selected by $b$-jet triggers differs from that of events selected by jet triggers, and each bucket in the mixed channel is scaled according to its corresponding luminosity. Consequently, the total integrated luminosity is $126~\ifb$ for the resolved channel, which relies exclusively on $b$-jet triggers, and ranges from $126$ to $136~\ifb$ for the mixed channel.

\subsection{Signal simulation and theoretical predictions}
\label{subsec:simulation}

The signal process is modeled using Monte Carlo (MC) simulation. Events are generated with \PYTHIA[8.306]~\cite{Bierlich:2022pfr} at leading-order in QCD, and \EVTGEN[1.7.0]~\cite{Lange:2001uf} is used for $B$-hadron decays. The NNPDF23LO~\cite{Ball:2012cx} parton density function is employed, together with the A14 underlying event tune~\cite{ATL-PHYS-PUB-2014-021}. The MC samples for the signal templates are produced using a fast simulation based on a parameterization of the calorimeter response~\cite{SIMU-2018-04}. The generation of the simulated event samples includes the effect of multiple \(pp\) interactions per bunch crossing, as well as contributions from interactions in neighboring bunch crossings (\pileup).

The signal is generated as a simplified narrow-width model, assuming a width of 10~\MeV for both $X$ and $S$. The $X$ boson is produced via gluon--gluon fusion (ggF) and decays exclusively into $SH \to 4b$. The $X$ mass spans the range $300~\GeV$--$6~\TeV$. The $S$ mass depends on $m_{\mathrm{X}}$, with a minimum value of $70~\GeV$ and a maximum of at least $130~\GeV$ below $m_{\mathrm{X}}$, ranging up to $5~\TeV$. The Higgs boson mass is set to $125~\GeV$. The region around $m_{\mathrm{S}} \approx 125~\GeV$ is not considered in this analysis, but it is covered by the search for resonant production of two Higgs bosons in the $4b$ final state~\cite{HDBS-2018-41}. The targeted two-dimensional mass grid, shown in Figure~\ref{fig:massgrid}, comprises 83 signal samples. For large mass splittings between $X$ and $S$, both the $H$ and $S$ bosons can be produced with large transverse momentum, giving rise to a fully boosted topology. This regime overlaps partially with the mixed channel coverage, but a dedicated analysis strategy would be required to exploit it fully.

\begin{figure}[t]
\center
\includegraphics[width=.6\textwidth]{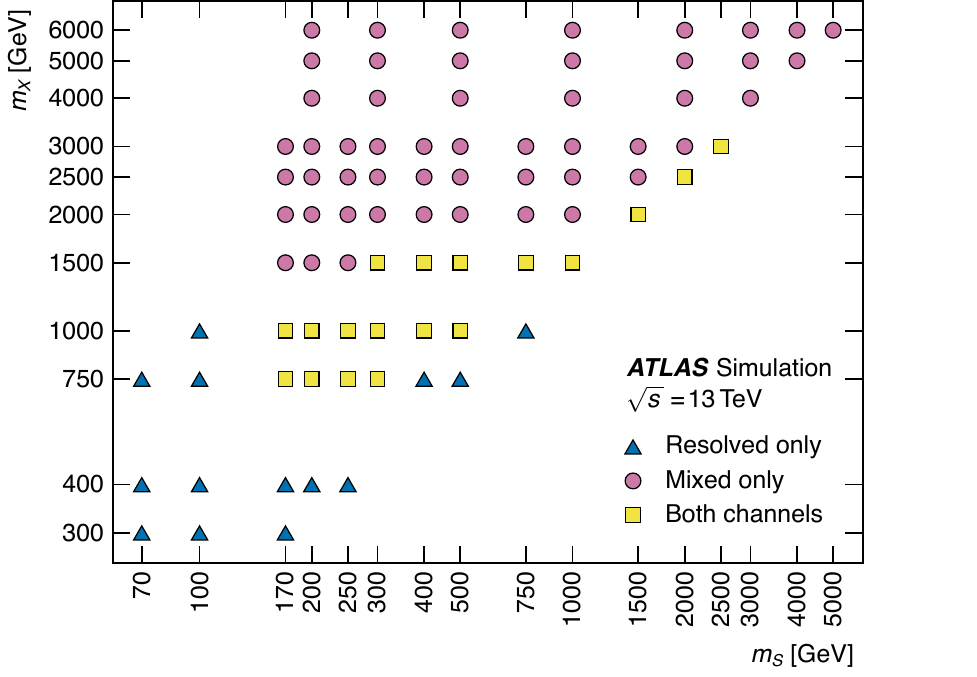}
\caption{The masses of the simulated signal samples in the $(m_{\mathrm{X}}, m_{\mathrm{S}})$ grid. Mass points probed by the resolved channel are shown with triangles, those probed by the mixed channel with circles, and those to which both channels are sensitive with squares.}
\label{fig:massgrid}
\end{figure}

The resolved and mixed channels probe complementary regions of the $(m_{\mathrm{X}}, m_{\mathrm{S}})$ plane. The mixed channel targets $m_{\mathrm{S}} > m_{\mathrm{H}}$, where the $H\to b\bar{b}$ decay products are sufficiently boosted to be captured by a single \largeR jet, but is limited at low $m_{\mathrm{X}}$ by the \largeR jet \pT requirements. The resolved channel instead targets configurations in which the \bquarks are more widely separated, but its sensitivity is suppressed when the $H$ and $S$ bosons are produced with a large separation in pseudorapidity, due to selection requirements. As a result, neither channel individually covers the full parameter space, and both are needed to achieve sensitivity across the explored mass grid, with overlapping coverage at intermediate $m_{\mathrm{X}}$ values. The resolved channel also probes mass points where excesses were previously reported by the CMS Collaboration.

For mass hypotheses covered by both the resolved and mixed analyses, the result from the channel with the best expected sensitivity is retained. Studies have shown that enforcing orthogonality between channels and combining the two analyses would provide only a marginal improvement in sensitivity.

The mass grid becomes progressively coarser as $m_{\mathrm{X}}$ and $m_{\mathrm{S}}$ increase. Due to detector effects, the resolution of the reconstructed resonances is significantly larger than the generated width. The spacing of the generated $m_{\mathrm{S}}$ values is generally smaller than the experimental resolution (typically 30\% relative to the $S$ mass), allowing a simple linear interpolation between neighboring mass hypotheses. In contrast, the mass resolution of the reconstructed $X$ candidate, $m_{\mathrm{X}}^{\mathrm{reco}}$, is typically less than 10\% relative to the $X$ mass and therefore narrower than the spacing between generated $m_{\mathrm{X}}$ values, preventing interpolation between neighboring $m_{\mathrm{X}}$ hypotheses.

TRSM predictions for cross-section $\sigma$ multiplied by branching ratio $\mathcal{B}\left(X\to SH\to 4b\right)$ in ggF production are used as a benchmark for $X$ masses up to 2500~\GeV. The heavy scalar production employs results from the NNLO+NNLL calculation~\cite{YR3} that accounts for top-quark mass induced effects up to NLO\@. For each fixed mass pair $(m_{\mathrm{X}}, m_{\mathrm{S}})$, the maximum allowed cross-section value is determined by scanning over the remaining free TRSM parameters (two vacuum expectation values and three mixing angles) and maximizing $\sigma \times \mathcal{B}\left(X \to SH \to 4b\right)$~\cite{trsmscans}. Parameter points incompatible with theoretical or experimental constraints~\cite{Bahl_2023} are discarded. If the predicted width of $X$ becomes larger than the experimental mass resolution, the parameter point is also excluded. The TRSM predictions are tabulated in Appendix~\ref{app:trsm}.

% End of text imported from the .//sections/samples.tex input file

% The next lines are included from the .//sections/objects.tex input file
\section{Physics objects}
\label{sec:objects}

Interaction vertices from the \pp collisions are reconstructed from at least two charged-particle tracks with $\pT>500~\MeV$ measured in the ID\@. The vertex with the largest sum of squared track momenta is selected as the hard-scatter primary vertex~\cite{PERF-2015-01}.

Small-radius (small-$R$) jets are reconstructed using the anti-$k_t$ algorithm with a radius parameter $R = 0.4$. The inputs to this algorithm are particle flow objects~\cite{PERF-2015-09}, which combine charged-particle tracks associated with the primary vertex with calorimeter energy deposits clustered using a topological algorithm~\cite{PERF-2014-07}. To avoid double-counting the energy of charged particles, the particle-flow algorithm subtracts from the topological clusters the energy deposits expected from particles associated with high-quality tracks and uses the track momentum measurements in their place. The inclusion of tracking information improves the jet energy resolution and helps ensure the reconstructed jets are associated with the primary vertex rather than \pileup interactions.

The momenta of reconstructed jets are calibrated in a multi-step procedure based on simulation, with additional corrections derived from data~\cite{JETM-2018-05}. Small-$R$ jets are required to satisfy $|\eta|<2.5$ and $\pT>40~\GeV (55~\GeV)$ in the resolved (mixed) channel, chosen to match the highest jet \pT threshold in the trigger, and must satisfy additional quality criteria to reject jets from non-collision backgrounds and calorimeter noise bursts~\cite{ATLAS-CONF-2015-029}. Jets with $\pT< 60~\GeV$ are further required to satisfy a selection based on the output of a neural-network-based jet vertex tagger algorithm (NNJVT)~\cite{PERF-2014-03}, which suppresses jets originating predominantly from \pileup interactions. Correction factors are applied to simulated events to account for differences in NNJVT efficiency between data and simulation. Jets must satisfy an $\eta$-independent selection that corresponds to efficiencies ranging from 80\% to 99\%, depending on the \pT.

To identify small-$R$ jets originating from \bquarks, the GN2 \btag algorithm~\cite{FTAG-2023-05} is employed. GN2 uses a transformer neural network to directly infer the jet flavor from low-level track and jet inputs, without relying on intermediate reconstruction of secondary vertices. The selected working point corresponds to a \bjet efficiency of 85\%, with rejection factors of 75 for light-flavor jets and 6.5 for $c$-jets, defined as the inverse of the corresponding misidentification probabilities, as evaluated for jets with $20 < \pT < 250~\GeV$ in simulated \ttbar events. Compared with the previously used DL1r algorithm~\cite{FTAG-2019-07}, employed for example in the $HH\to b\bar{b}b\bar{b}$ analysis~\cite{HDBS-2018-41}, GN2 provides a dramatically improved performance. At the same \bjet efficiency, the light- (charm-) jet rejection is enhanced by factors of 1.9 (2.2)~\cite{FTAG-2019-02,FTAG-2020-08,FTAG-2018-01}.

In addition, the mixed channel employs anti-$k_t$ \largeR jets with a radius parameter $R = 1.0$ to reconstruct boosted $H\to b\bar{b}$ decays. These jets are built from Unified Flow Objects (UFOs)~\cite{JETM-2018-06}, which extend the particle-flow approach by optimally combining tracking and calorimeter information, including inputs from Track-Calo Clusters~\cite{ATL-PHYS-PUB-2017-015}, to enhance jet substructure reconstruction. To ensure jet constituents originate from the primary vertex and to remove soft radiation, a sequence of grooming techniques is applied. The jet constituents are first corrected using constituent subtraction and the SoftKiller algorithm~\cite{JETM-2018-06}, followed by soft-drop grooming~\cite{Larkoski_2014} to remove residual soft-QCD contributions. The reconstructed jet energy and mass are calibrated through simulation-based corrections that restore the response to the particle level on average~\cite{PERF-2015-09}, with additional data-driven corrections applied to refine the energy scale and resolution~\cite{JETM-2018-02}. \LargeR jets used in the mixed channel are required to satisfy $\pT>250~\GeV$ and $|\eta|<2.0$.

To identify \largeR jets originating from $H\to b\bar{b}$ decays, the GN2X \btag algorithm~\cite{ATL-PHYS-PUB-2023-021} is employed. Like GN2, the GN2X algorithm uses a transformer neural network that is trained on track inputs and high-level information about each \largeR jet. GN2X implements a discriminant constructed from the probabilities for the jet to originate from $H \to b\bbar$, $H \to c\cbar$, top-quark, and QCD processes, with coefficients to weight each of the background contributions.  The coefficient for top-quark processes is set to zero to mitigate an observed dependence of the respective probability on the data-taking period.
Working points are defined such that the multijet background mistag efficiency is constant as a function of the \largeR jet mass, as derived with Pythia8 multijet events~\cite{ATL-PHYS-PUB-2019-017}. This working point definition reduces correlations between the \largeR jet mass and the multijet background rejection, improving the data-driven background estimate based on mass sidebands. A working point for the 0.37\% multijet mistag efficiency corresponds to a $H\to b\bar{b}$ tagging efficiency of 55\%. Signal efficiency corrections are measured in $Z\to b\bar{b}+\text{jets}$ events~\cite{ATL-PHYS-PUB-2021-035}.
GN2X provides a very significantly improved performance compared with the previous approach, in which \btag was applied to track-jets ghost-associated with the \largeR jets~\cite{Cacciari_2008}, providing an approximately twofold decrease in the multijet and top background rejection for a 55\% $H\to b\bar{b}$ efficiency~\cite{ATL-PHYS-PUB-2023-021}.

% End of text imported from the .//sections/objects.tex input file

% The next lines are included from the .//sections/selection.tex input file
\section{Event selection}
\label{sec:selection}

The data analysis was performed using the analysis framework documented in Ref.~\cite{easyjet}, developed within the ATLAS Collaboration for collision data analysis. Two analysis channels are defined to target the resolved and mixed topologies defined in \cref{sec:intro}, as described below.

\subsection{Resolved channel}
Events in the resolved channel are required to pass either a  \texttt{2b+1j} trigger or a \texttt{2b+2j} trigger, as described in \cref{sec:samples}, and to contain at least four jets with $\pT >40$~\GeV and $|\eta| <  2.5$ satisfying the GN2 \btag requirement. The $S$ and $H$ candidates are reconstructed by pairing the four highest-\pT \btagged jets using a BDT, implemented with the Gradient Boost algorithm in TMVA~\cite{TMVA}. The BDT is trained on kinematic variables from all possible jet pairings to identify the combination that best matches the underlying decay topology.

Due to the variety of kinematics across the $(m_{\mathrm{X}}, m_{\mathrm{S}})$ grid, the signal samples in the resolved channel are divided into two training groups. When the mass difference between the $X$ and $S$ resonances is close to $m_H=125$~\GeV, both intermediate bosons are produced nearly at rest, leading to a back-to-back topology of the $b$-jets from each scalar decay. These scenarios define the compressed group, while all other resolved signal points are assigned to the non-compressed group.

The input variables include separation in pseudorapidity ($\Delta\eta$) and azimuthal angle ($\Delta\phi$) between the jets in each pair, their angular separation ($\Delta R$), the transverse momentum of each pair, the invariant mass of the four \bjets, and the difference between the reconstructed $S$ and $H$ candidate masses ($m_{\mathrm{S}}^{\mathrm{reco}}$ and $m_{\mathrm{H}}^{\mathrm{reco}}$, respectively). To prevent biasing events towards the reconstructed mass distributions, the invariant masses of the \bjet pairs are not included among the BDT input variables. The invariant mass of the four \btagged jets is identical across all pairing hypotheses and therefore carries no discriminating power on its own, but it parametrizes the pairing decision, allowing the BDT to learn the $m_{\mathrm{X}}$-dependent correlations between the angular variables and the correct pairing.

Dijet pairs coming from a resonance with mass $m$ with transverse momentum \pt have an opening angle described by $\Delta R_{jj} \sim 2 m / \pt$.
Therefore, after the \bjet pairs are formed using the BDT, the pair with larger angular separation is identified as the more massive boson.
The mass of the dijet pairs is not used directly to assign the $S^\mathrm{reco}$ and $H^\mathrm{reco}$ candidates to avoid sculpting the $m_\mathrm{H}^\mathrm{reco}$ distribution used to define the control region for the background estimate.
The pseudorapidity separation between the reconstructed $S$ and $H$ candidates, $\Delta \eta_{\mathrm{SH}}^\mathrm{reco}$, is required to satisfy $|\Delta \eta_{\mathrm{SH}}^\mathrm{reco}| < 1.5$, to reduce multijet backgrounds. This selection exploits the more central signature of signal events compared with the background.
The efficiency of the BDT-based jet pairing combined with the scalar assignments ranges from 57--99\%. Efficiency is generally higher for more boosted topologies and in the compressed regime due to the dedicated training.

The reconstructed Higgs candidate mass, $m_{\mathrm{H}}^\mathrm{reco}$, is used to define the regions employed in the background estimate. The region $107.8 < m_{\mathrm{H}}^\mathrm{reco} < 148.8~\GeV$, centered on the Higgs boson mass was blinded during background estimation, while the adjacent sideband regions, $59.3 < m_{\mathrm{H}}^\mathrm{reco} < 107.8~\GeV$ and $148.8 < m_{\mathrm{H}}^\mathrm{reco} < 190.7~\GeV$, serve as control regions (CRs) in which the background model is trained. Background estimates are defined inclusively for several $m_\mathrm{S}$ hypotheses: the resolved channel has three background estimate regions, as visualized in \cref{fig:BDT_kinematic_regions}, corresponding to the non-compressed $m_{\mathrm{S}} > m_{\mathrm{H}}$, non-compressed $m_{\mathrm{S}} < m_{\mathrm{H}}$, and compressed regimes (all compressed signal hypotheses have $m_{\mathrm{S}} > m_{\mathrm{H}}$).

After the background estimate is obtained, the signal purity is increased by defining a signal region (SR) for each $m_\mathrm{S}$ hypothesis in the $(m_\mathrm{H}^\mathrm{reco}, m_\mathrm{S}^\mathrm{reco})$ plane. Events are required to satisfy $X_\mathrm{SH}^\mathrm{reco} < 1.6$, where $X_\mathrm{SH}^\mathrm{reco}$ is a distance variable defined as:
\begin{equation}
\label{eq:xsh}
X_\mathrm{SH}^\mathrm{reco} = \sqrt{\left(\frac{m_\mathrm{H}^\mathrm{reco} - m_H'}{r_{f,H} \cdot m_\mathrm{H}^\mathrm{reco}}\right)^2 +
\left(\frac{m_\mathrm{S}^\mathrm{reco} - m_S'}{r_{f,S} \cdot m_\mathrm{S}^\mathrm{reco}}\right)^2},
\end{equation}

where $m_H'$ and $m_S'$ denote the center of the SR along the $m_\mathrm{H}^\mathrm{reco}$ and $m_\mathrm{S}^\mathrm{reco}$ axes, respectively. The parameter $m_H'$ is fixed to $125~\GeV$. The parameter $m_S'$ is determined from simulation for each $m_\mathrm{S}$ hypothesis: it is defined as the median of the reconstructed $m_\mathrm{S}^\mathrm{reco}$ distribution of signal events passing the $4b$ selection, averaged over the simulated samples with the same generated $m_\mathrm{S}$ and different $m_\mathrm{X}$ values. The parameters $r_{f,H}$ and $r_{f,S}$, referred to as resolution factors, determine the width of the SR along the $m_\mathrm{H}^\mathrm{reco}$ and $m_\mathrm{S}^\mathrm{reco}$ directions, respectively. The value $r_{f,H} = 0.1$ is fixed, while $r_{f,S}$ is computed from the same simulated $m_\mathrm{S}^\mathrm{reco}$ distributions as half the ratio of the width of the central 68\% interval to the median. This $X_\mathrm{SH}^\mathrm{reco}$ definition was inspired by the signal region used for the ATLAS $HH\rightarrow 4b$ analysis~\cite{HDBS-2018-41}, where the resolution factors allow the width of the SR to increase at lower reconstructed resonance masses where the dijet mass resolution degrades. The SR definitions depend only on the $m_\mathrm{S}$ hypothesis and are independent of $m_\mathrm{X}$; each SR is subsequently evaluated for several $m_\mathrm{X}$ signal hypotheses in the statistical fit. The data in the $(m_\mathrm{H}^\mathrm{reco}, m_\mathrm{S}^\mathrm{reco})$ plane, the CRs, and the $X_\mathrm{SH}^\mathrm{reco} < 1.6$ SRs are illustrated in \cref{fig:BDT_kinematic_regions}.

\begin{figure}[t]
\centering
\subfloat[]{
\includegraphics[width=0.49\textwidth]{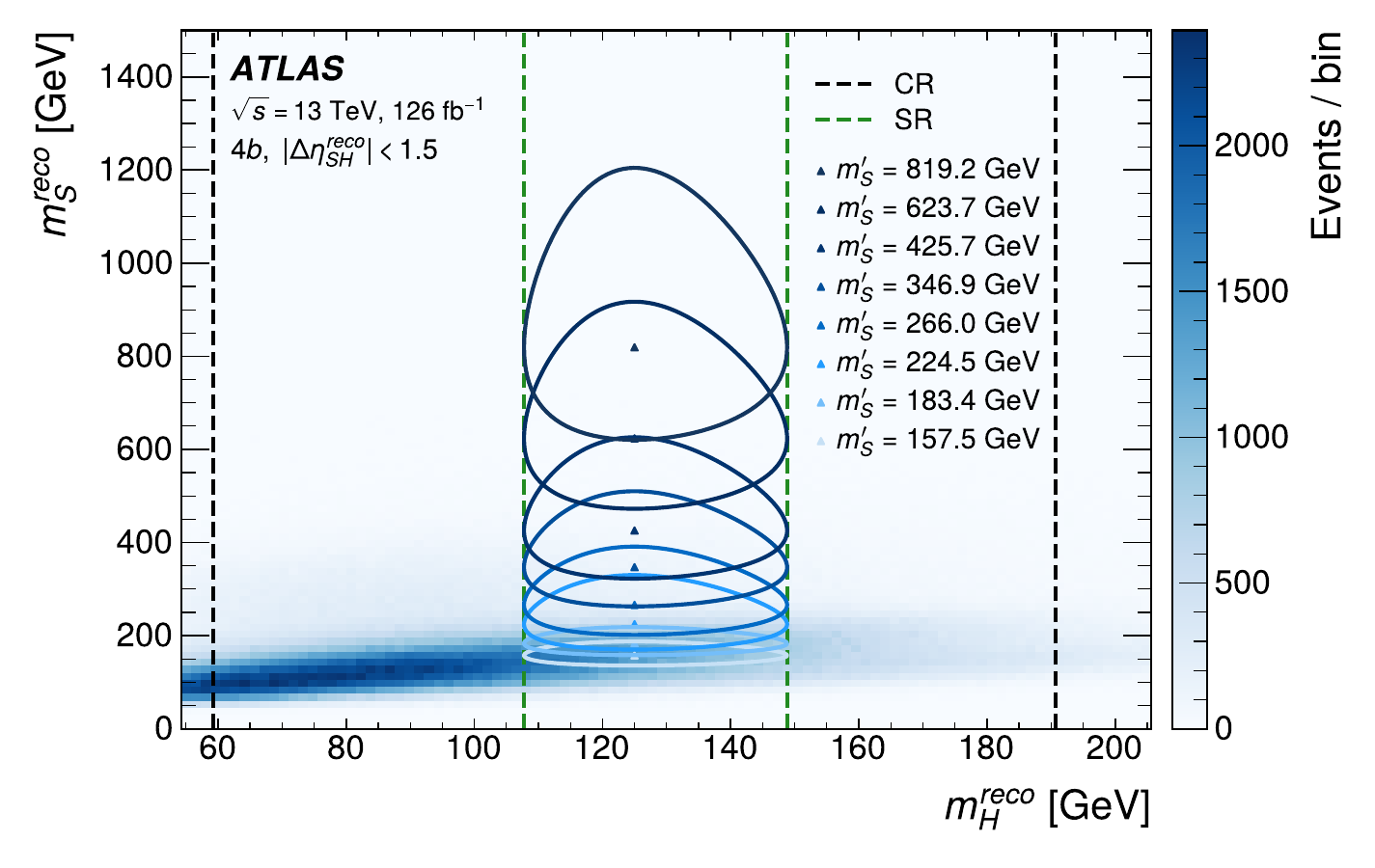}
\label{fig:BDT_mass_plane1}
}
\subfloat[]{
\includegraphics[width= 0.49\textwidth]{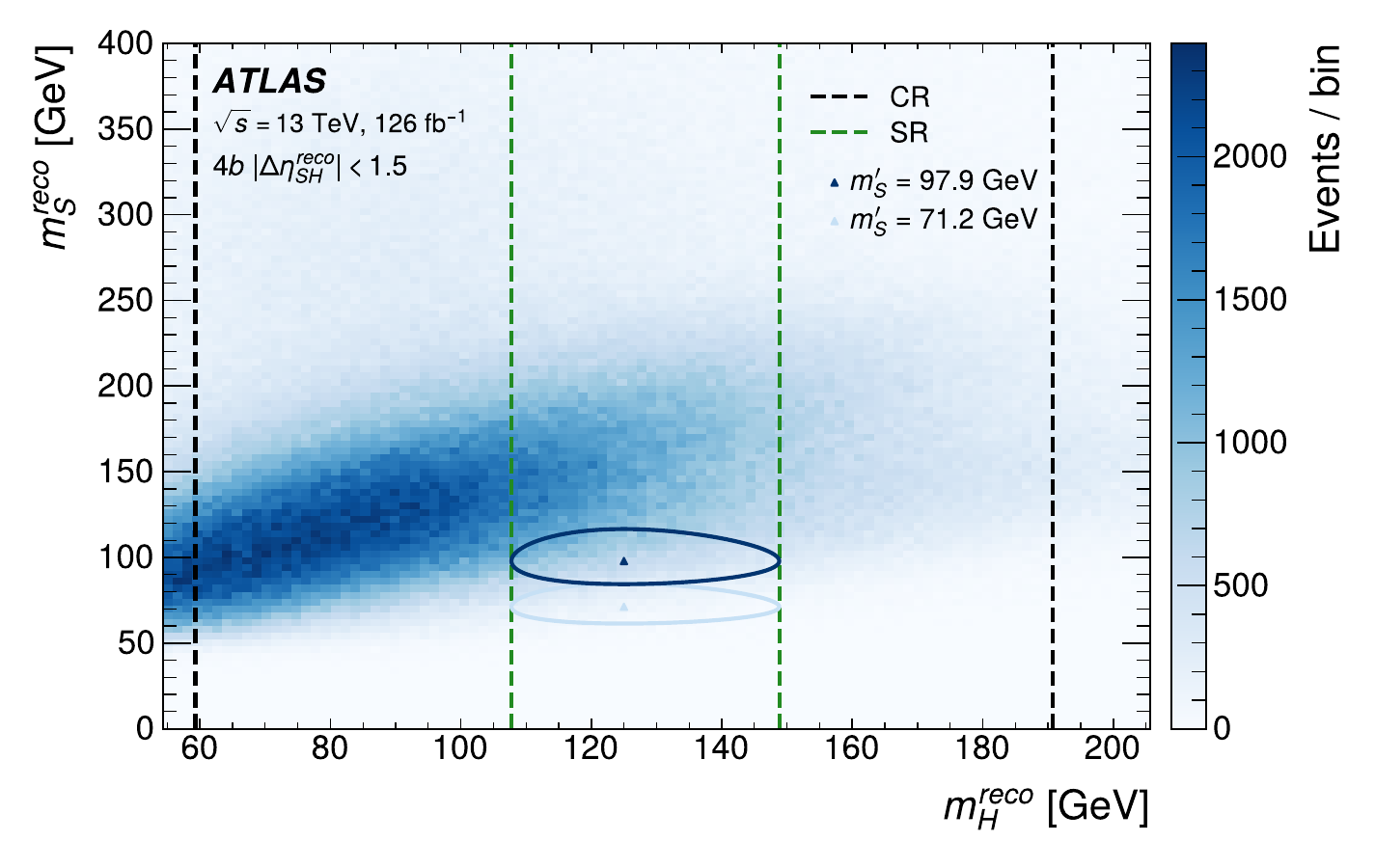}
\label{fig:BDT_mass_plane2}
}\\
\subfloat[]{
\includegraphics[width= 0.49\textwidth]{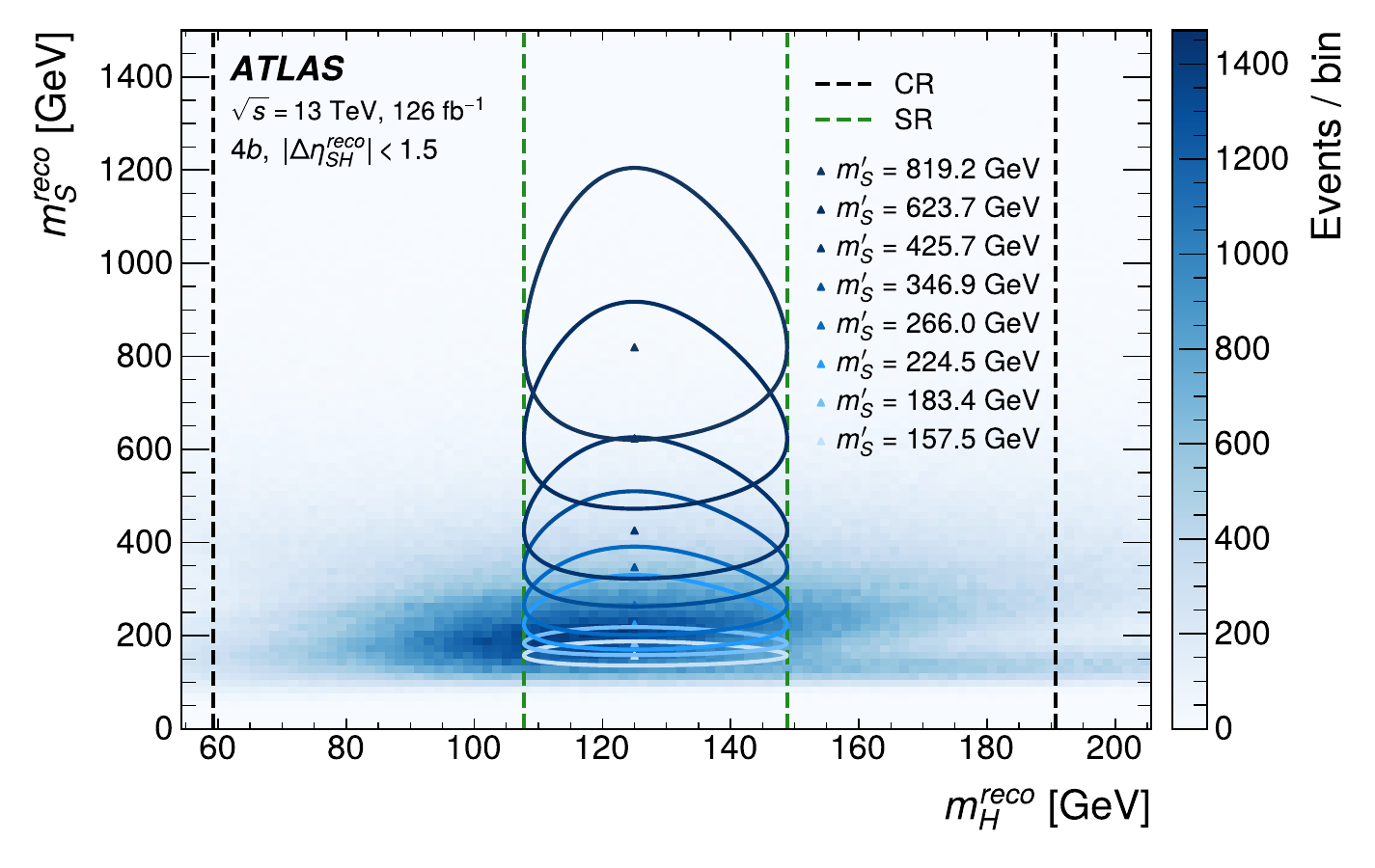}
\label{fig:BDT_mass_plane3}
}
\caption{
Data for the resolved analysis for different BDT training groups and different cases of scalar mass assignment, represented in the reconstructed $(m_\mathrm{H}^\mathrm{reco}, m_\mathrm{S}^\mathrm{reco})$ plane, for (a) the non-compressed regime with $m_{\mathrm{S}} > m_{\mathrm{H}}$, (b) the non-compressed regime with $m_{\mathrm{S}} < m_{\mathrm{H}}$, and (c) the compressed regime with $m_{\mathrm{S}} > m_{\mathrm{H}}$. The black dashed lines indicate the CR (the outer pair, used for training the background model), and the green dashed lines indicate the SR (the inner pair, blinded for the training). The rounded convex shapes indicate the final cut $X_\mathrm{SH}^\mathrm{reco} < 1.6$ applied after deriving the background model, with a different $X_{SH}^\mathrm{reco}$ definition for each $m_\mathrm{S}$ hypotheses. These regions are independent of $m_\mathrm{X}$. Each resulting $X_\mathrm{SH}^\mathrm{reco}$ SR is subsequently evaluated for several $m_\mathrm{X}$ signal hypotheses in the statistical fit.
}
\label{fig:BDT_kinematic_regions}
\end{figure}

To further enhance the sensitivity of the resolved channel, the SR defined by $X_\mathrm{SH}^\mathrm{reco} < 1.6$ is partitioned into six exclusive categories for the fit, based on the variables $X_\mathrm{SH}^\mathrm{reco}$ and $|\Delta\eta_\mathrm{SH}^\mathrm{reco}|$. Two intervals in $X_\mathrm{SH}^\mathrm{reco}$,  $\left[ 0, 0.95 \right)$ and $\left[ 0.95, 1.6 \right]$, are combined with three intervals in $|\Delta\eta_\mathrm{SH}^\mathrm{reco}|$, $\left[ 0, 0.5 \right)$, $\left[ 0.5, 1 \right)$, and $\left[ 1, 1.5 \right]$, resulting in six categories.

A validation region (VR), denoted $3b$,  is defined to assess the performance of the data-driven background model. It is constructed by modifying the nominal \btag selection to require exactly three jets satisfying the GN2 \btag criterion, with the fourth jet taken as the highest-\pT jet failing that requirement. Pseudo-control and pseudo-signal regions are defined within this region following the same procedure as in the nominal selection. The kinematic similarity of the $3b$ region to the nominal $4b$ selection makes it well-suited for this purpose, and any residual discrepancies observed are propagated as a non-closure uncertainty (quantified as the relative difference between the $3b$ data and the $3b$ prediction) in the background prediction.

\subsection{Mixed channel}
The mixed channel targets events in which the Higgs boson is highly Lorentz-boosted. Events are required to pass either a \largeR jet trigger or a 2b+1j trigger, as described in \cref{sec:samples}. Following the trigger selection, events must contain at least two small-$R$ jets with $\pT> \SI{55}{\GeV}$ and $|\eta| < 2.5$, and at least one \largeR with $\pT > \SI{250}{\GeV}$, $|\eta| < 2.0$, and $70 < m < \SI{200}{\GeV}$.

The $H$ candidate is reconstructed as the highest-$\pt$ large-$R$ jet and is required to pass the GN2X tagger requirement.  After the $H$ candidate is identified, overlapping small-$R$ jets are removed, where overlap is defined as an angular separation $\Delta R < 1.0$ between the $H$ candidate and the small-$R$ jet.

The $S$ candidate is reconstructed from the two highest-$\pT$ small-$R$ jets that remain, both of which are required to satisfy the 85\% working point of the GN2 tagger. A requirement of $m_\mathrm{S}^\mathrm{reco} > \SI{70}{\GeV}$ on the $S$ candidate is applied in the nominal selection, as the mixed channel considers only signal hypotheses with $m_{\mathrm{S}} \geq \SI{170}{\GeV}$. Once the $H$ and $S$ candidates are identified, events are assigned to trigger buckets based on the $\pT$ of the $H$ candidate, as described in \cref{sec:samples}. Events failing the corresponding trigger requirement for their assigned bucket are rejected.

Similarly to the resolved $3b$ region, a VR denoted \emph{1b} is defined to test the background modeling and derive a non-closure uncertainty. In this region, exactly one of the two small-$R$ jets forming the $S$ candidate must satisfy the 85\% GN2 working point, while the other must fail it.

The SR is defined by $110 < m_{\mathrm{H}}^\mathrm{reco} < 140~\GeV$. The $m_{\mathrm{H}}^\mathrm{reco}$ sideband regions, defined by $70 < m_{\mathrm{H}}^\mathrm{reco} < 110~\GeV$ and $140 < m_{\mathrm{H}}^\mathrm{reco} < 200~\GeV$, constitute the CR used to derive the background prediction, as described in \cref{sec:background}.

For mass pairs with $m_{\mathrm{X}} \le 3~\TeV$ and $m_{\mathrm{S}} \le 2~\TeV$, distinct signal regions are defined for each truth $m_{\mathrm{S}}$ value by imposing additional cuts on $m_{\mathrm{S}}^\mathrm{reco}$. The allowed $m_{\mathrm{S}}^\mathrm{reco}$ range is chosen as the 99\% probability interval of a skew-normal function fit to the reconstructed $m_{\mathrm{S}}$ distribution of all selected signal events with the same truth $m_{\mathrm{S}}$.  For the ($m_{\mathrm{X}}, m_{\mathrm{S}}) = (3~\TeV, 2.5~\TeV)$ mass pair and all mass pairs with $m_{\mathrm{X}} \ge 4~\TeV$, additional cuts are not applied because they would not improve the sensitivity.

% End of text imported from the .//sections/selection.tex input file

%

% The next lines are included from the .//sections/background.tex input file
\section{Background estimation}
\label{sec:background}

\begin{figure}[]
\centering
\includegraphics[width=\textwidth,trim={0 200 0 280},clip]{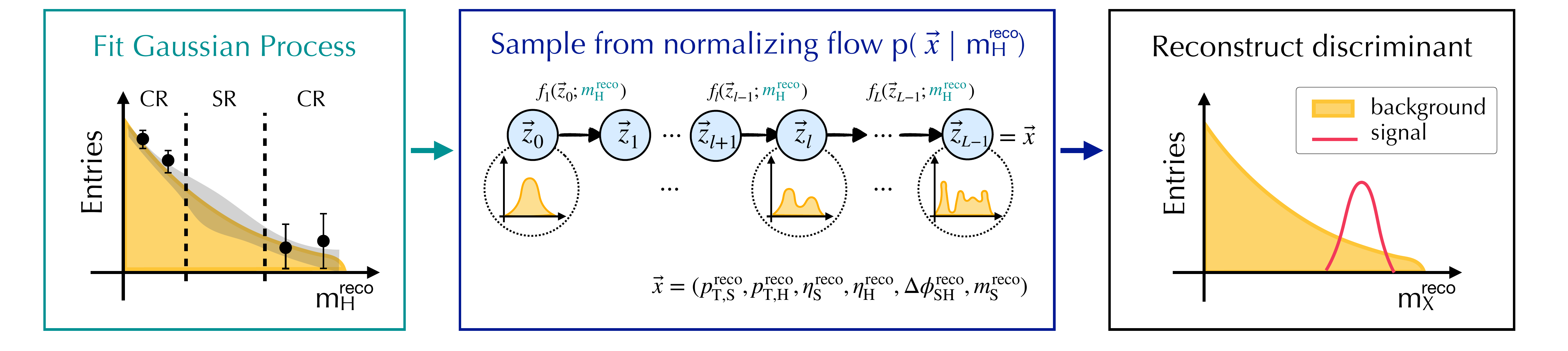}
\caption{Overview of the background model.
(1) A Gaussian Process (GP) is fitted to the $m_{\mathrm{H}}$ distribution in the CR to obtain the background prediction and its uncertainty.
(2) A normalizing flow, conditioned on the GP prediction in the signal region (SR), is used to generate the remaining kinematics.
(3) Reconstructed four-vectors are used to compute the final discriminant. The SR definition depends on the $m_{\mathrm{S}}$ hypothesis.}
\label{fig:bkg-pred-graphic}
\end{figure}

This analysis performs a data-driven background estimate via a high-dimensional interpolation, illustrated by Figure~\ref{fig:bkg-pred-graphic}. The background in the SR is modeled by the joint density $p(\vec{x}, m_{\mathrm{H}}^\mathrm{reco})$, in which $\vec{x}$ encodes the kinematic properties of the $H$ and $S$ candidates.
This density is decomposed as, $p(\vec{x}, m_{\mathrm{H}}^\mathrm{reco})=p_{\mathrm{GP}}(m_\mathrm{H}^\mathrm{reco})\cdot p_{\mathrm{flow}}(\vec{x}|m_{\mathrm{H}}^\mathrm{reco})$.
The Gaussian process (GP) is used to predict the shape $p_\mathrm{GP}(m_\mathrm{H}^\mathrm{reco})$ as well as the overall normalization.
The $p_\mathrm{flow}$ term is learned with a normalizing flow, a universal approximator of continuous probability distributions, which interpolates between the CR and the SR through the conditioning variable, $m_{\mathrm{H}}^\mathrm{reco}$.
The final discriminant, $m_\mathrm{X}^\mathrm{reco}$, is obtained by sampling from the combined GP and flow predictions.
The validity of the interpolation constitutes the key assumption of the method; dedicated uncertainties are therefore derived to account for potential non-closure effects (Section~\ref{subsec:data-driven-systematics}). Further details are provided in the following subsections.

This model jointly predicts the dominant background contributions: QCD multijet events, which account for approximately 90\% of the total background, and $t\bar{t}$ production. Subleading contributions from $Z\to b\bar{b}$ and from single Higgs boson decays into $b\bar{b}$, with the latter produced via ggF, vector-boson fusion, and in association with a $Z$ boson, are also considered in this data-driven estimate. Studies with MC simulation show that these processes contribute only at the percent level and their resonant shape does not introduce any localized bias in the background estimate, hence their overall impact is negligible.

\subsection{Gaussian processes}
\label{subsec:gps}

A GP is a Bayesian non-parametric method that generalizes the idea of a multivariate Gaussian to describe a distribution over functions~\cite{Rasmussen}. A GP defines a prior with the covariance structure specified by a kernel function.  A commonly used kernel, and the one adopted in this work, is the squared exponential kernel $K(m,m') = \exp\left(- \frac{(m-m')^2}{2 \ell^2}\right)$, where the length scale $\ell$ governs the smoothness of the predicted functions. For any finite collection of points, the GP is described by a multivariate Gaussian. As Gaussians are analytically tractable, conditioning on the training data yields a predictive posterior for the test data with an uncertainty that grows with distance from the training points, suitable both for interpolation and extrapolation. The only hyperparameter is the length scale $\ell$, which is determined by maximizing the log likelihood of the training data. This approach was employed successfully in several ATLAS analyses for smoothing, interpolation, and background modeling~\cite{HDBS-2018-41, HIGG-2019-13, ATL-PHYS-PUB-2020-028, HIGG-2021-06, HIGG-2019-23, HMBS-2024-28_v1}. Here, the GP is also used for an interpolation, and the novelty of this application is that the GP prediction is used to condition a normalizing flow, as described in the next section.

A GP is fit using \textsc{scikit-learn}~\cite{scikit-learn} to the $m_{\mathrm{H}}^\mathrm{reco}$ histogram in each CR.\@ %
The training input is the $m_{\mathrm{H}}^\mathrm{reco}$ bin centers (abscissas) and corresponding yields (ordinates), and their Poisson uncertainties. Inputs are scaled to the range $[0,1]$, while targets are standardized to have zero mean and unit variance. To account for differences between trigger
configurations across data-taking periods, a separate GP is trained for each year.

\cref{fig:mH_prediction} shows the $m_{\mathrm{H}}^\mathrm{reco}$ distributions together with the GP predictive posterior in the signal region for the resolved non-compressed ($m_{\mathrm{S}} < m_{\mathrm{H}}$) channel (\cref{fig:mH_4b_deta_noncompressed_minS_18}) and the mixed channel (\cref{fig:mH_4b_nom_mixed_18}).

As noted in \cref{sec:selection}, in the resolved channel, the assignment of dijet pairs to $S$ or $H$ depends on the jet opening angle. Since the CR definition depends on this assignment, separate background estimates are constructed for $m_{\mathrm{S}} < m_{\mathrm{H}}$ and $m_{\mathrm{S}} > m_{\mathrm{H}}$. The compressed reconstruction also requires an independent background prediction.

The GP background model provides a smooth interpolation into the SR, with uncertainties that increase moderately within the interpolated region and in CR bins with fewer training events.
The solid histogram represents the mean of the GP posterior for the $m_{\mathrm{H}}^\mathrm{reco}$ histogram with discrete $(m_{\mathrm{H}}^{(i)}, y^{(i)})$ points describing the $i$ bin centers and yields, respectively.
This histogram is converted into $m_{\mathrm{H}}^\mathrm{reco}$ samples by first sampling $m_{\mathrm{H}}^{(i)}$ bin centers using inverse transform sampling~\cite{inv-xfrom}.
These discrete $m_{\mathrm{H}}^{(i)}$ samples are then smeared uniformly within each bin width to obtain a continuous $m_{\mathrm{H}}^\mathrm{reco}$ distribution, which is used to condition the normalizing flow prediction in the SR.\@
The total predicted SR yield is normalized using the relation
$n_\mathrm{SR} = n_\mathrm{CR}^\mathrm{obs} \cdot n_\mathrm{SR}^\mathrm{pred} / n_\mathrm{CR}^\mathrm{pred}$.

\begin{figure}[t]
\centering
\subfloat[]{
\includegraphics[height=6cm]{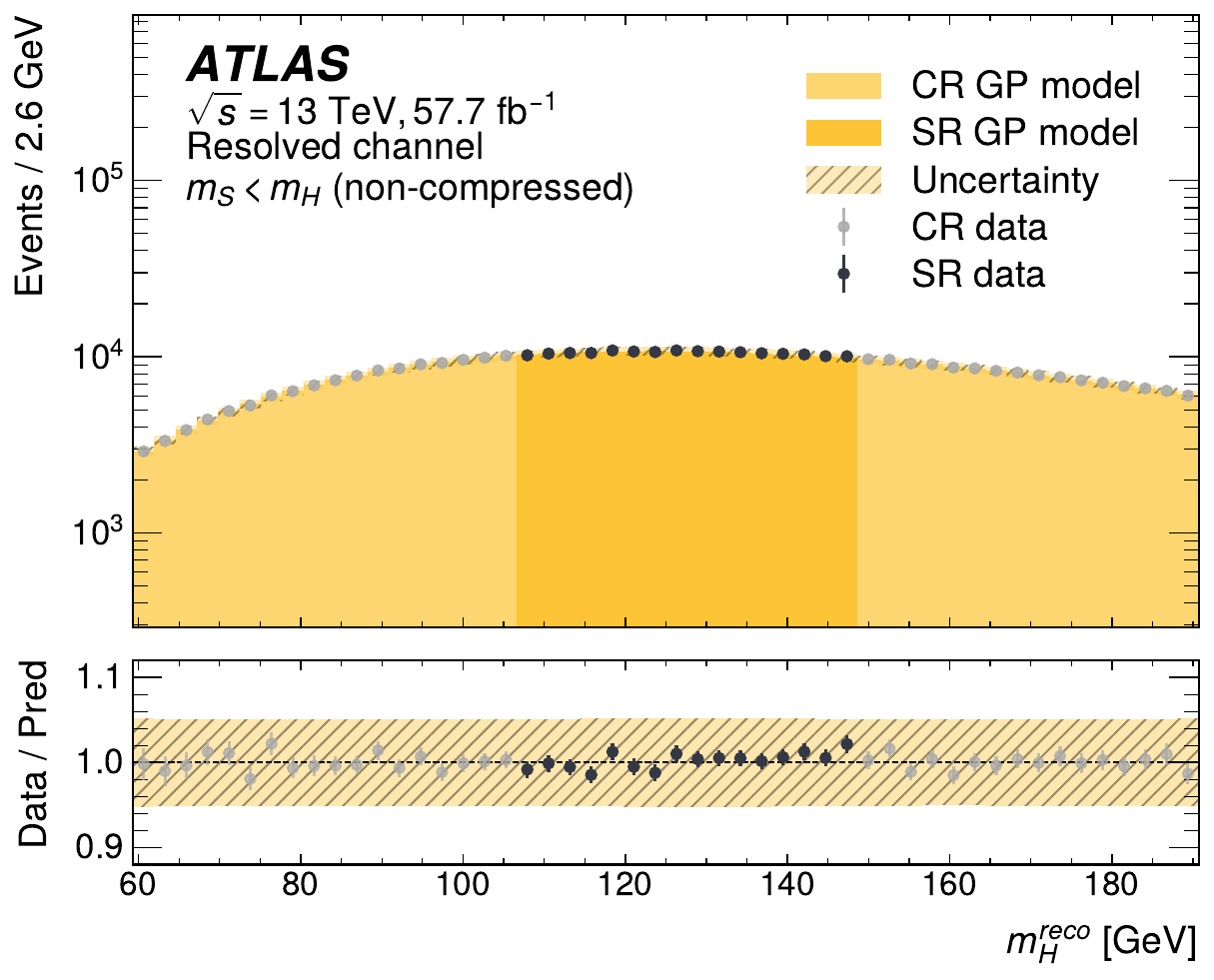}
\label{fig:mH_4b_deta_noncompressed_minS_18}
}
\subfloat[]{
\includegraphics[height=6cm]{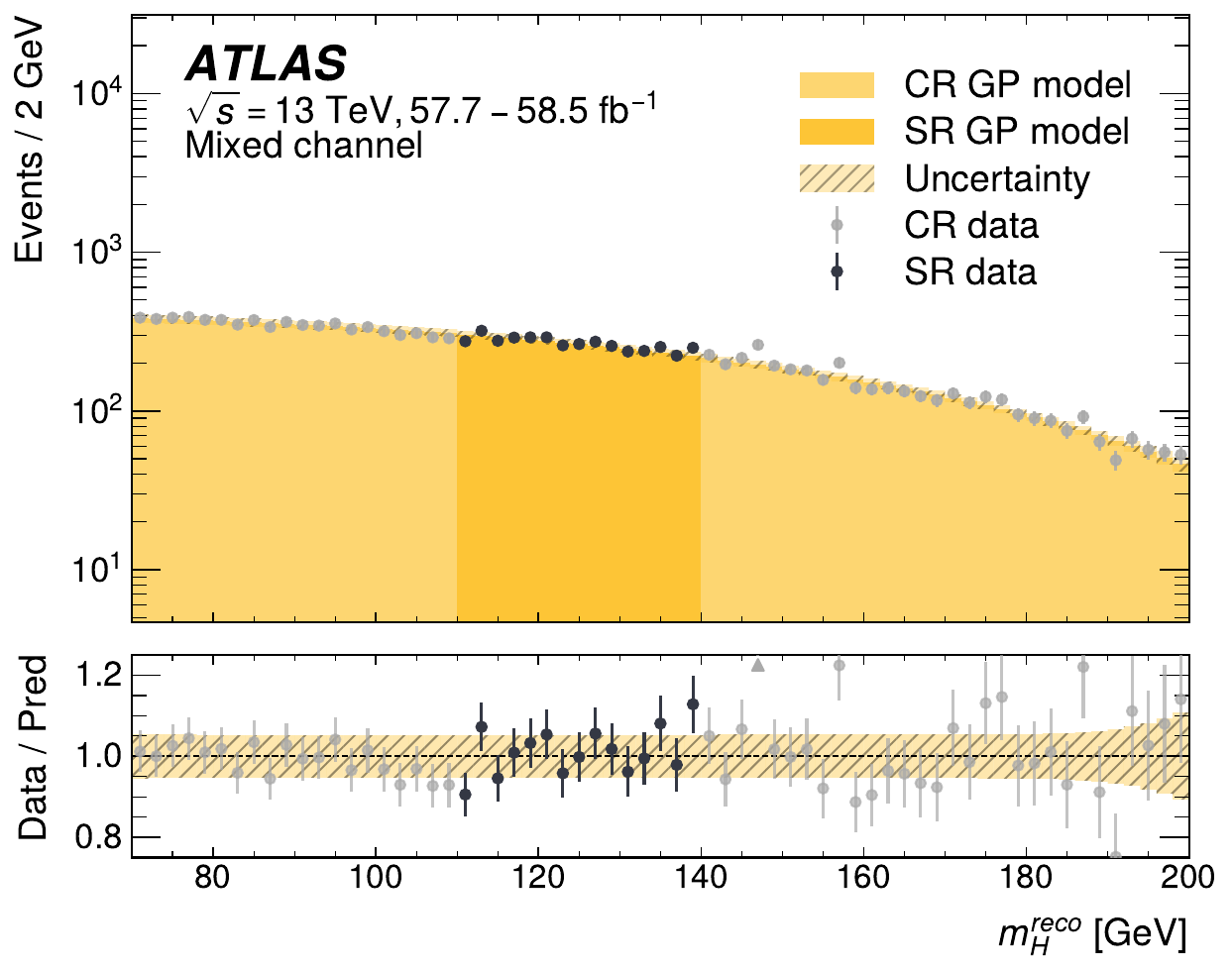}
\label{fig:mH_4b_nom_mixed_18}
}
\caption{The GP fit to the $m_{\mathrm{H}}^\mathrm{reco}$ distribution, for (a) the resolved non-compressed $m_{\mathrm{S}} < m_{\mathrm{H}}$ background estimate, and (b) the mixed background estimate. The lighter colors indicate the CR included in the fit, while the darker colors indicate the SR not seen during the training. The hatched areas indicate the prefit uncertainties from the GP model. Predictions for the 2018 data are shown for both topologies. The lower panel shows the ratio of observed data to the background prediction. A small arrow indicates the case when the ratio value exceeds the plot range.}
\label{fig:mH_prediction}
\end{figure}

\subsection{Normalizing flows}
A normalizing flow is a generative model that transforms a known base density, such as a Gaussian distribution, into a target density by a sequence of invertible transformations, denoted $f_i$~\cite{1505.05770}. Each transformation is required to be invertible and have a tractable Jacobian determinant.

The forward mode of the flow, $f = f_1 \circ \cdots \circ f_L$, maps samples from the base distribution to the target space, as illustrated in \cref{fig:bkg-pred-graphic}.
The inverse transformation, $f^{-1}$, enables evaluation of the probability density via the change-of-variables formula

\begin{equation}
p_x \left(\vec{x}\right) = p_z(\vec{z}) \left| \frac{dz}{dx} \right| = p_z \left(f^{-1}\left(\vec{x}\right)\right) \left| \frac{df^{-1}}{dx} \right|,
\end{equation}
where $p_z\left(\vec{z}\right)$ is the Gaussian base density and $\vec{x}$ represents the features to be modeled.
The model parameters are optimized by maximizing the likelihood of the training data in the CR, or equivalently by minimizing the negative log-likelihood,

\begin{equation}
\mathcal{L} = - \frac{1}{N} \sum_{\vec{x} \sim \mathrm{CR}} \log p(\vec{x} \mid m_{\mathrm{H}}^\mathrm{reco}, \mathrm{year}).
\end{equation}

A variety of flow architectures were developed to maximize expressivity while maintaining computational tractability~\cite{1910.13233, 1912.02762}.
Rational quadratic neural spline flows (RQ-NSF)~\cite{1906.04032} is a widely used architecture due to its flexibility and computational efficiency.
In this work, the RQ-NSF architecture is used to model the conditional probability distribution of the background event kinematics. The transformations are parameterized by splines, whose parameters are predicted by neural networks. Each flow is conditioned on $m_{\mathrm{H}}$ and the data-taking year, to enable a smooth interpolation from the CR to the SR and to allow a single model to be trained on data from all three years.

The flows predict the kinematics of the $S$ and $H$ candidates $\vec{x} = (p_{\mathrm{T},\mathrm{S}}^\mathrm{reco}, p_{\mathrm{T},\mathrm{H}}^\mathrm{reco}, \eta_{\mathrm{S}}^\mathrm{reco}, \eta_{\mathrm{H}}^\mathrm{reco}, \Delta \phi_{\mathrm{SH}}^\mathrm{reco}, m_{\mathrm{S}}^\mathrm{reco})$. To improve modeling performance, variables with sharp features are transformed before training: logarithmic transformations are applied to $p_{\mathrm{T},\mathrm{H}}^\mathrm{reco}$, $p_{\mathrm{T},\mathrm{S}}^\mathrm{reco}$, and $m_{\mathrm{S}}^\mathrm{reco}$, while the variable $\log(\pi - \Delta \phi_{\mathrm{SH}}^\mathrm{reco})$ is used for the azimuthal separation. All inputs are standardized to zero mean and unit variance.
While for the final discriminant, it is possible to model fewer variables with the flow (e.g., this resolved discriminant would only need to minimally model $\vec{x} = (m_X^\mathrm{reco},\Delta \eta_{SH}^\mathrm{reco}, m_\mathrm{S}^\mathrm{reco})$, it was empirically found that modeling the scalar's four-momentum $\vec{x}$ gives better closure in signal-like validation regions than modeling the high-level variables $\vec{x}'$ directly.

The model architecture is chosen to balance expressivity and robustness against overfitting. The model consists of ten transformation layers, with splines defined using $K=4$ bins.
The spline parameters are predicted by ResNets~\cite{resnet} with 32 hidden units. To avoid overfitting, a dropout fraction~\cite{dropout} of 10\% was used and an L2 regularization auxiliary loss term, scaled by $10^{-6}$. The model is trained using the Adam optimizer~\cite{kingma2017adammethodstochasticoptimization}  with a learning rate of $10^{-3}$.

A common approach to estimating the uncertainty associated with neural network predictions is to train an ensemble of models and quantify the uncertainty from the variance of their predictions~\cite{deepensembles}.
This accounts for the uncertainty associated with the stochastic optimization process, as different trainings can converge to distinct local minima~\cite{fort2020deepensembleslosslandscape}.
Each flow is trained 25 times, with the mean of the predictions taken as the central value and the spread used to define the associated uncertainty, as described in Section~\ref{subsec:data-driven-systematics}.

The GP posterior carries an intrinsic uncertainty, which is propagated by sampling 25 histogram realizations of $m_{\mathrm{H}}^\mathrm{reco}$ from it. Each sampled histogram is paired with an independent flow training, allowing the GP uncertainty to propagate through to the final discriminant, though in practice, the flow training uncertainty dominates.

The prediction in the SR samples from $p_\mathrm{flow}(\vec{x}|m_\mathrm{H}^\mathrm{reco})$, conditioning on the $m_{\mathrm{H}}^\mathrm{reco}$ samples from the GP\@. The flow prediction $\vec{x}$ along with the GP $m_{\mathrm{H}}^\mathrm{reco}$ samples is combined to reconstruct the four-vectors of the $S$ and $H$ candidates. From these, derived observables such as $m_{\mathrm{X}}^\mathrm{reco}$ and $\Delta \eta_{\mathrm{SH}}^\mathrm{reco}$ are computed.
The performance of the normalizing flow in modeling representative kinematic distributions is illustrated in \cref{fig:flow_pred_mixed}, which compares SR data to background model predictions for selected observables in the mixed channel.

\begin{figure}[t]
\centering
\subfloat[]{
\includegraphics[height=6cm]{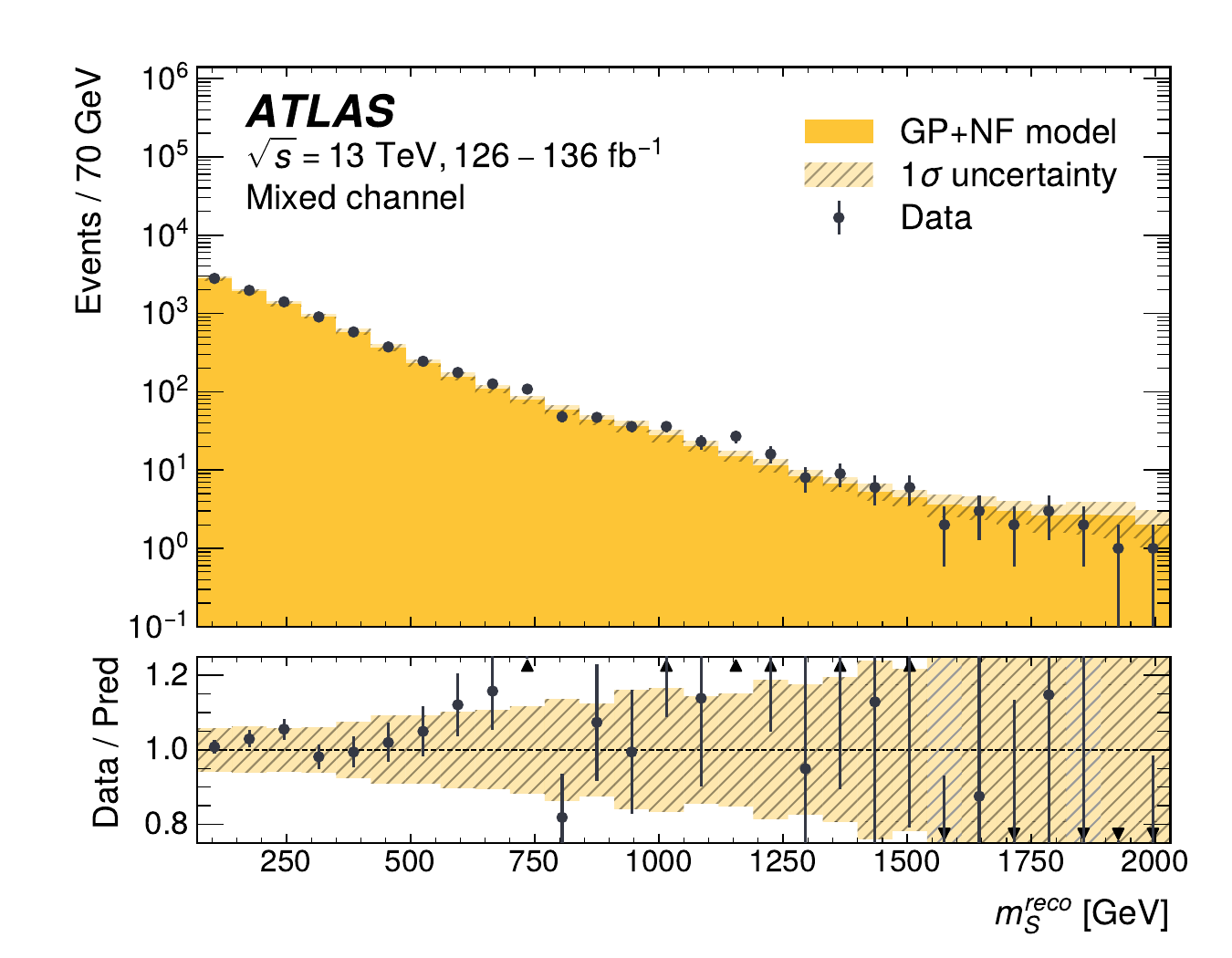}
\label{fig:mS_4b_nom_mixed_18}
}
\subfloat[]{
\includegraphics[height=6cm]{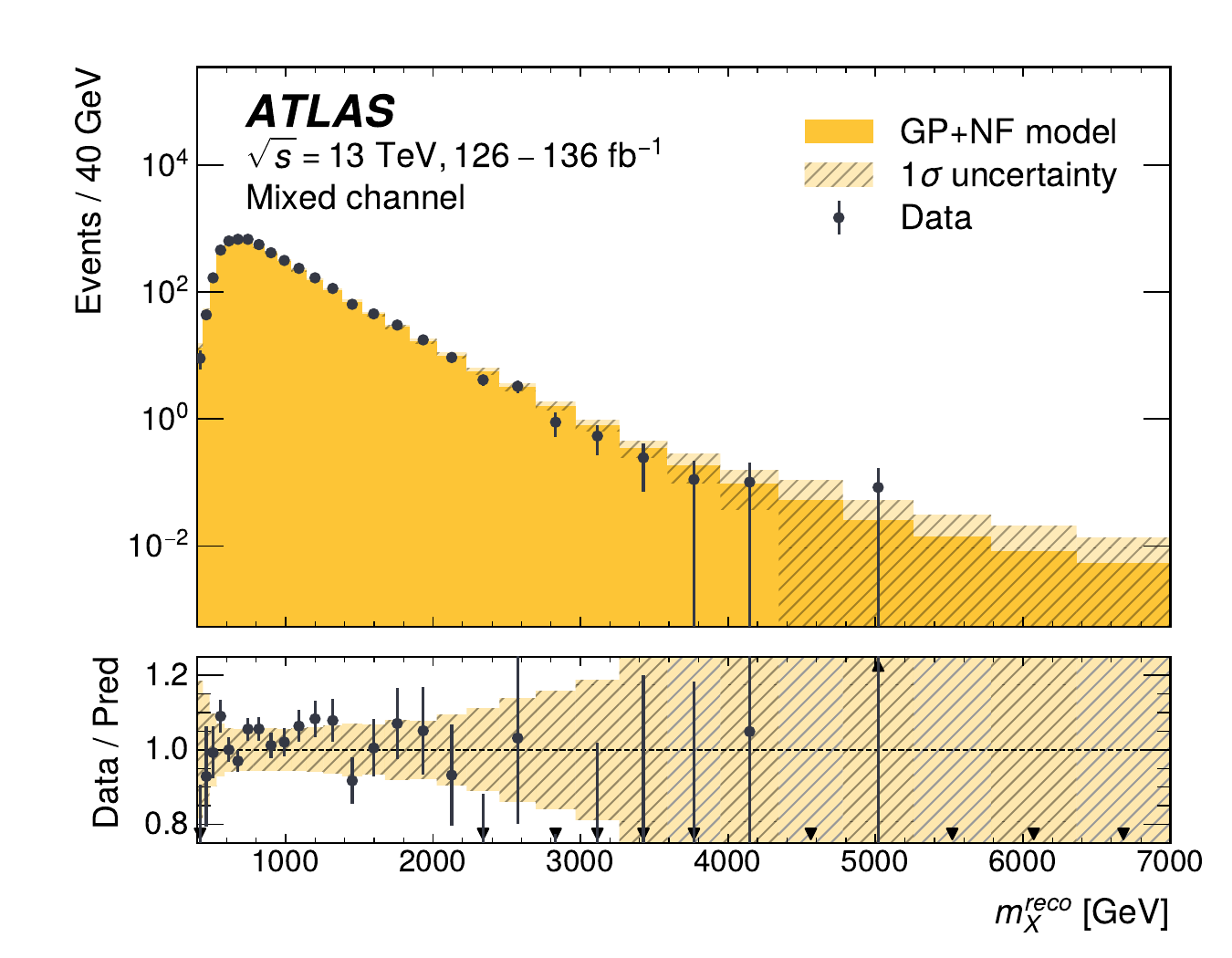}
\label{fig:mX_4b_nom_mixed_18}
}
\caption{Selected kinematic distributions predicted by the background model for the SR of the mixed channel, compared with the real data. (a) shows the $m_{\mathrm{S}}^\mathrm{reco}$ distribution, and (b) displays the final discriminant $m_{\mathrm{X}}^\mathrm{reco}$. Both quantities are reconstructed from the GP and flow predictions. The hatched areas indicate the prefit uncertainties from the full background model. The lower panel shows the ratio of observed data to the background prediction. A small arrow indicates the case when the ratio value exceeds the plot range.}
\label{fig:flow_pred_mixed}
\end{figure}

Each resolved background estimate is trained in the $m_H$ CR after applying the $|\Delta \eta_{\mathrm{SH}}^\mathrm{reco}| < 1.5$ requirement. After training the three resolved background estimates, for each $m_\mathrm{S}$ hypothesis, the corresponding $X_\mathrm{SH}^\mathrm{reco} < 1.6$ requirement is then applied to the generated events. In the mixed analysis, the background prediction is trained inclusively in $m_\mathrm{S}^\mathrm{reco}$, and then the $m_\mathrm{S}^\mathrm{reco}$ selection is subsequently applied for each $(m_{\mathrm{X}}, m_{\mathrm{S}})$  mass hypothesis.

Compared with previous approaches targeting a similar final state, where the CR definition constitutes a dominant source of uncertainty~\cite{HDBS-2019-29}, the interpolation-based strategy adopted here reduces the non-closure uncertainty by approximately a factor of two. To the extent that the interpolation is not perfect, residual mismodeling is quantified through non-closure studies in dedicated validation regions, which are defined orthogonally to the signal regions and used to validate the procedure before unblinding.

% End of text imported from the .//sections/background.tex input file

% The next lines are included from the .//sections/systematics.tex input file
\section{Systematic uncertainties}
\label{sec:systematics}

Systematic uncertainties are grouped into three main categories: experimental, theoretical, and background modeling (data-driven).
Experimental uncertainties account for potential differences between the detector response in data and simulation, and are applied to all quantities derived from simulated samples. Theoretical uncertainties arise from the modeling of simulated processes. Since the background estimate is fully data-driven, experimental and theoretical uncertainties affect only the signal prediction. Background uncertainties are associated with the modeling of the $m_{\mathrm{X}}^\mathrm{reco}$ distribution and its extrapolation to the signal region.

\subsection{Experimental and theoretical uncertainties}

Experimental uncertainties include contributions from the integrated luminosity measurement, \pileup modeling, trigger efficiencies, calibration of the GN2 $b$-tagger and the GN2X tagger, and the reconstruction and calibration of jets. These uncertainties affect both the normalization and shape of the signal templates and are propagated to the statistical analysis.

The uncertainty in the integrated luminosity of the combined data from 2016, 2017, and 2018 is 0.84\%, using the same methodology as in Ref.~\cite{DAPR-2021-01}, and obtained using the LUCID-2 detector~\cite{LUCID2} for the primary luminosity measurements, complemented by measurements using the inner detector and calorimeters. %

Uncertainties related to \pileup modeling arise from the reweighting of simulated events to reproduce the \pileup conditions observed in data~\cite{STDM-2015-05}.

Jet-related uncertainties include contributions from the calibration and resolution of both small-\ensuremath{R} and \largeR jets. For small-\ensuremath{R} jets, these comprise uncertainties in the jet energy scale and jet energy resolution, while for \largeR jets they additionally include uncertainties in the jet mass scale and jet mass resolution. These uncertainties are derived using simulation-based techniques and \emph{in situ} measurements~\cite{JETM-2018-05,PERF-2014-03,JETM-2023-02} and are propagated to the signal yields.

Uncertainties associated with flavor tagging are evaluated for the GN2 and GN2X algorithms and account for uncertainties in the calibration of the \bjet~\cite{FTAG-2023-05,FTAG-2018-01,FTAG-2020-08} and $H \to b\bar{b}$ jet tagging efficiencies~\cite{ATL-PHYS-PUB-2021-035}.

Additional uncertainties are assigned to account for differences between the fast simulation used to produce the signal samples and the full \textsc{Geant4} simulation~\cite{Agostinelli:2002hh}.
The prescriptions for this uncertainty are centrally provided by the ATLAS simulation group.
These include dedicated variations affecting the response of small-\ensuremath{R} and \largeR jets, as well as the GN2 calibration, and are propagated to the corresponding signal templates.

Additional uncertainties are associated with the modeling of trigger efficiencies, including the jet triggers and $b$-jet triggers used in the event selection~\cite{TRIG-2018-08}. The jet $E_\mathrm{T}$ dependence is derived in a di-leptonic $t\bar{t}$ selection and propagated to the signal prediction. With a preliminary online \bjet calibration, checks for a few representative signal samples demonstrated that a 20\% signal normalization uncertainty conservatively accounts for the difference in the online $b$-tagging efficiency between data and MC.

A theoretical uncertainty associated with the modeling of the signal process is included and treated as a normalization uncertainty. A conservative uncertainty of 5\% is assigned to the signal yield in both the resolved and mixed channels~\cite{HDBS-2018-41}. This uncertainty covers potential variations arising from the parton shower, underlying event modeling, hadronization~\cite{ATL-PHYS-PUB-2014-021}, and parton distribution functions.

\subsection{Data-driven background uncertainties}
\label{subsec:data-driven-systematics}

As described in \cref{sec:background}, the background prediction is obtained from an ensemble of 25 independently trained models in each CR, each initialized with a different random seed. Each background estimate is taken as the mean prediction of the ensemble.
The spread of the ensemble predictions covers the uncertainty from the network initialization and optimization (network retraining) and the finite number of events in the CR~\cite{deepensembles}.
Shape uncertainties used in the fit are derived from this spread.
This approach has been used in other particle physics analyses, including searches for $HH \rightarrow 4b$~\cite{HDBS-2018-41,HDBS-2019-29,CMS-HIG-24-010} and neural simulation-based inference in the off-shell Higgs boson analysis~\cite{SOFT-2023-01,HIGP-2024-14}.

For the resolved analysis, the 25 background prediction templates are decomposed using principal component analysis (PCA) to identify the dominant directions of variation induced by the different initial seeds and residual statistics-dominated components. The eigenvectors of the PCA decomposition define the shape variations used in the fit, scaled by the square root of the corresponding eigenvalues. %
The number of retained correlated components is determined using the autocorrelation\footnote{Autocorrelation measures the degree to which a binned variation pattern is correlated with itself under shifts in the bin index. Large autocorrelation indicates a coherent shape variation, while small or rapidly vanishing autocorrelation indicates bin-to-bin statistical fluctuations.} of the corresponding PCA templates. The remaining components are dominated by statistical uncertainties from the finite data in the CR and are added in quadrature. They are then incorporated as uncorrelated bin-by-bin uncertainties in the final fit. An autocorrelation value of 0.6 is used to make the distinction.

\begin{figure}[h!tb]
\centering
\includegraphics[width=\textwidth]{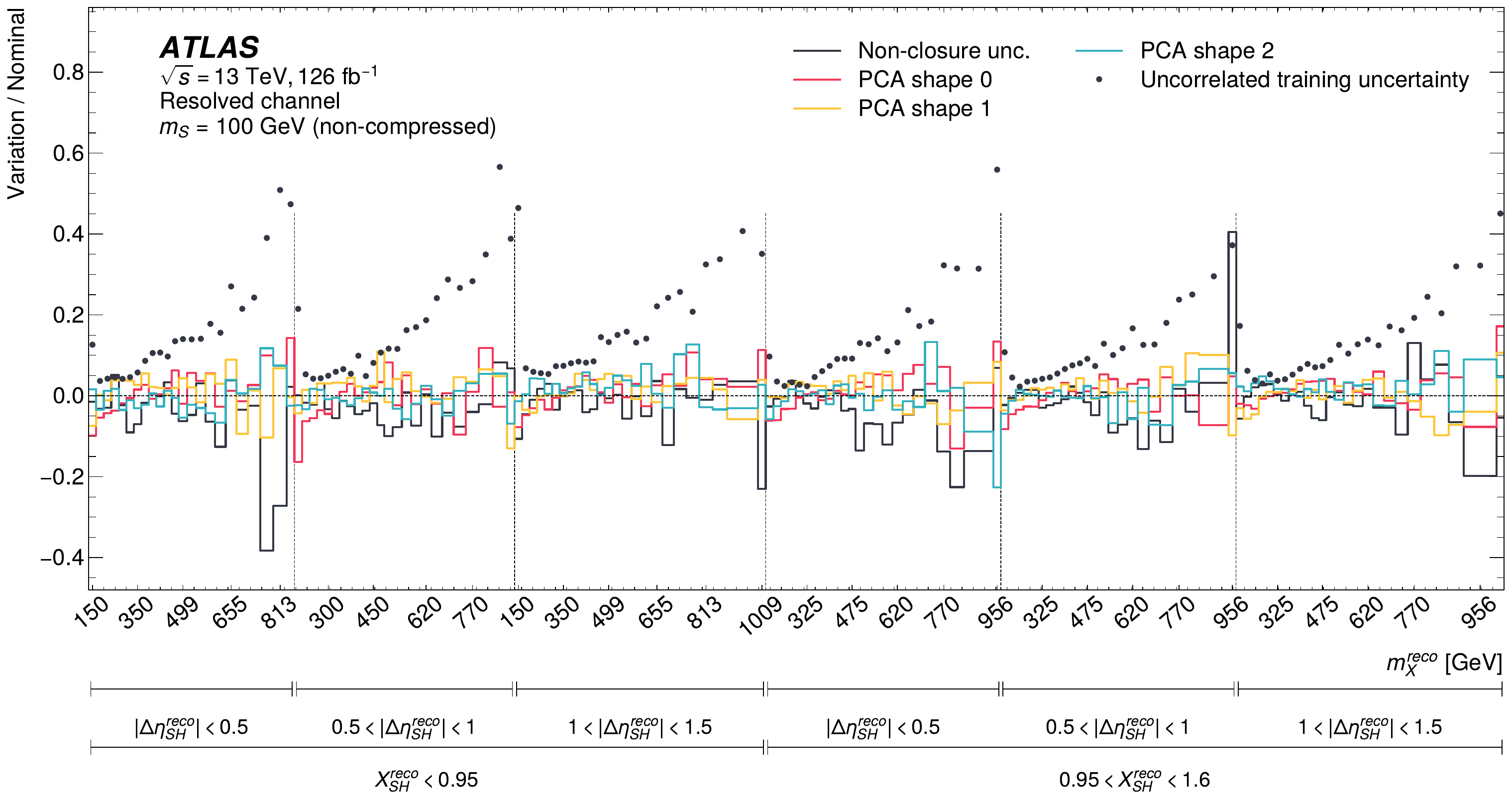}
\hfill
\caption{Relative variations of the background uncertainties with respect to the nominal background prediction for the resolved non-compressed $m_{\mathrm{S}} = 100$~\GeV SR\@. The figure shows the non-closure uncertainty, the correlated shape uncertainties derived from the principal-component decomposition of the ensemble predictions, and the residual uncorrelated training uncertainty.}
\label{fig:sys_vali_closure:100_minS_nc}
\end{figure}

For the mixed analysis, which relies on a simpler prediction model extracted in a single category, the shape uncertainty is derived directly from the ensemble spread. For each of the 25 models, a relative variation is constructed from the deviation of that model with respect to the nominal prediction, scaled by the inverse square root of the number of models. These variations are implemented as correlated per-bin shape uncertainties, with each model contributing one independent nuisance parameter.

An additional non-closure uncertainty is assigned to account for residual differences between the observed and predicted $m_{\mathrm{X}}^\mathrm{reco}$ distributions.  This uncertainty is evaluated using the $3b$ and $1b$ VRs described in \cref{sec:selection}, which are kinematically similar but orthogonal to the corresponding SRs. The uncertainty is assigned to account for residual differences between the observed data and the predicted background distributions of the discriminating variable $m_{\mathrm{X}}^\mathrm{reco}$. It is computed on a bin-by-bin basis from the difference between the observed and predicted yields in the validation region. The resulting variation is then propagated to the signal region by scaling it by the ratio of the expected yields in the signal and validation regions.

Finally, an additional $5\%$ normalization uncertainty is assigned to the background prediction to account for potential mismodeling of the overall background normalization obtained from the GP fits.

Examples of the resulting data-driven background systematic templates for the resolved non-compressed $m_{\mathrm{S}} = 100~\GeV$ signal region and the mixed $m_{\mathrm{S}}=\text{750}~\GeV$ signal region are shown in \cref{fig:sys_vali_closure:100_minS_nc} and \cref{fig:mixed_bkg_syst_750}, respectively.

\begin{figure}[t]
\centering
\includegraphics[width=0.5\textwidth]{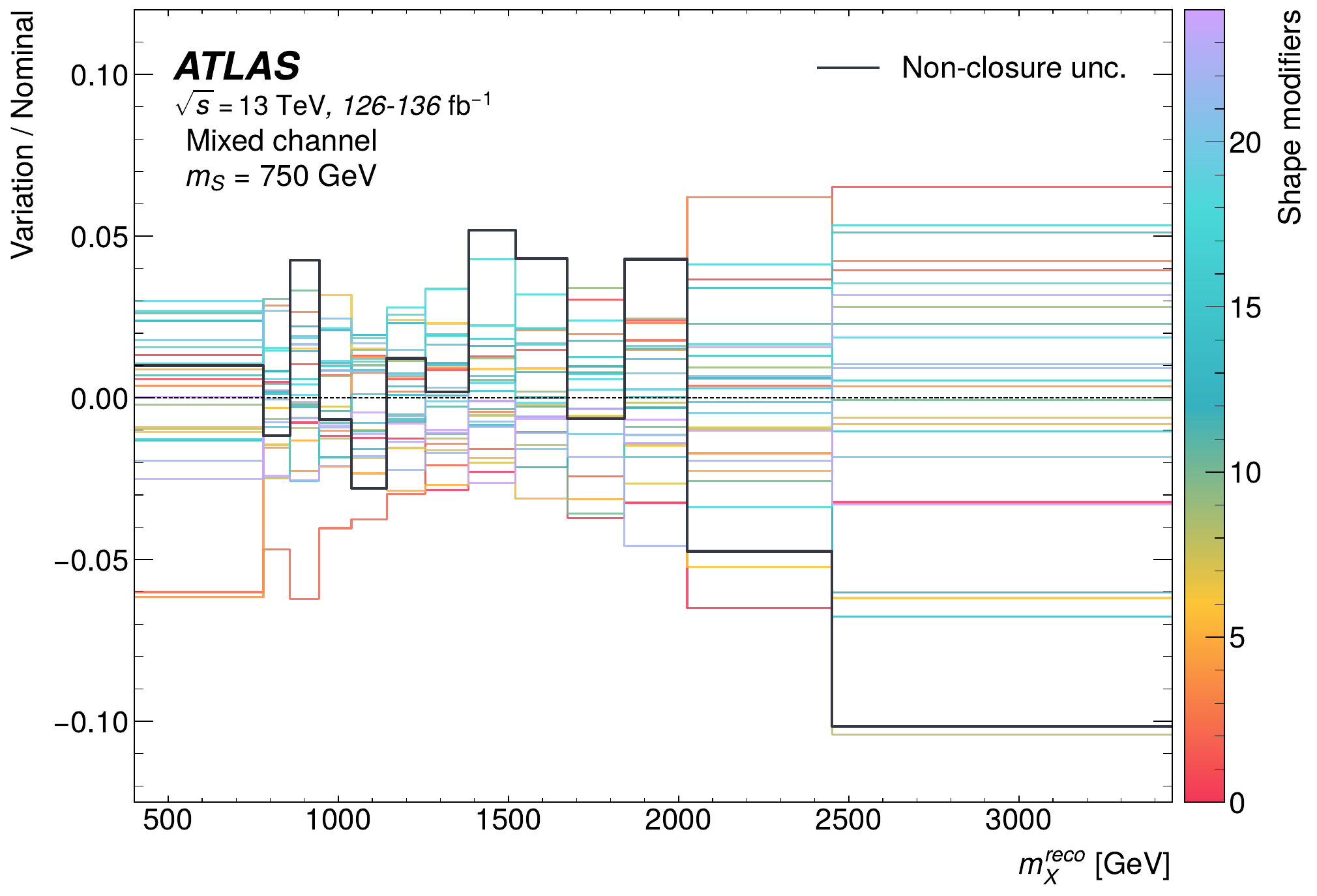}
\hfill
\caption{Relative variations of the data-driven background uncertainties with respect to the nominal background prediction for the mixed $m_{\mathrm{S}} = 750$~\GeV SR\@. The figure shows the non-closure uncertainty together with the correlated shape uncertainties derived from the ensemble of background predictions.}
\label{fig:mixed_bkg_syst_750}
\end{figure}

\cref{tab:systematic_impact} summarizes the impact of the different groups of systematic uncertainties in the expected upper limits. The impact of each group is evaluated as the absolute difference between the expected upper limit obtained with all systematic uncertainties included and the corresponding limit obtained after excluding that group of uncertainties. The quoted ranges correspond to the minimum and maximum percentage changes in the expected upper limits observed across the mass hypotheses relevant for each analysis channel.

Uncertainties associated with the data-driven background modeling constitute the dominant contribution in the resolved channel, modifying the expected limits by 44--73\%, depending on the signal mass hypothesis. In the mixed channel, background modeling uncertainties remain among the leading contributions, modifying the expected limits by up to 26\%, while flavor-tagging uncertainties associated with the GN2X calibration also have a significant impact for many signal mass points, reaching up to 32\%. Smaller contributions arise from uncertainties in jet reconstruction and trigger efficiencies. %

\begin{table}
\centering
\caption{Summary of the relative impact of each source of uncertainty on the analysis sensitivity.
The relative impact of an uncertainty source reflects how much the expected 95\% CL upper cross-section limit would improve, as a percentage, if that uncertainty were zero. For the data statistical uncertainty, the impact is assessed relative to the ideal limit of zero, which would be obtained with infinite data statistics, and is therefore given by the stat-only limit divided by the limit with all systematic uncertainties included. For the systematic uncertainties, it is assessed as the absolute difference between the limit with all systematic uncertainties included and the limit with that source excluded, divided by the limit with all systematic uncertainties included. The interval stated for each impact corresponds to the range of mass points in each channel.}
\label{tab:systematic_impact}
\begin{tabular}{lcc} \toprule
\multirow{2}{*}{Uncertainty source} 				& \multicolumn{2}{c}{Relative impact of uncertainties [\%]} \\ \cline{2-3}
& Resolved Channel		& Mixed Channel	\\ \toprule
\textbf{Data statistical uncertainty}				& 24--51				& 55--74			\\ \midrule
\textbf{All systematic uncertainties}		& 49--76				& 26--45			\\ \midrule
\hspace{1em}\textbf{Experimental}				& 48--76				& 26--45			\\ \midrule
\hspace{2em}\textbf{Detector response}		& 9--15				& 24--33			\\
\hspace{3em}Luminosity and \pileup			& <1 				& <1				\\
\hspace{3em}Flavor tagging				& 0--5.4				& 22--32			\\
\hspace{3em}Jet reconstruction				& 0.1--2.7				& 0.4--2.5			\\
\hspace{3em}Trigger efficiency				& 8--9				& <1			 	\\
\hspace{2em}\textbf{Background modeling}		& 44--73				& 1--26			\\
\hspace{3em}Network retraining and CR stats	& 36--63				& 0.1--23			\\
\hspace{3em}Non-closure and normalization	& 0--25				& 0.4--7.2			 \\ \midrule
\hspace{1em}\textbf{Theoretical}				& <1					& <1				 \\ \midrule
\hspace{1em}\textbf{MC statistical}				& <1 				& <1				 \\ \midrule
\end{tabular}
\end{table}

\subsection{Validation regions}
\label{subsec:validation-regions}

The background prediction and its associated uncertainties are validated in dedicated VRs.
Each analysis regime employs a VR designed to be kinematically similar to the corresponding signal region while remaining statistically independent. In the resolved analysis, the VR is defined by inverting the requirement on the pseudorapidity separation between the $S$ and $H$ candidates, $\Delta \eta_{\mathrm{SH}}^\mathrm{reco}$.  In the mixed analysis, the VR is defined by requiring the \largeR jet associated with the $H$ candidate to fail the GN2X 0.37\% mistag-rate working point while passing the looser 1.5\% working point, our `GN2X loose' region. This region is used solely to validate the background modeling and the associated background uncertainty prescription. Since the background prediction is fully data-driven, no additional GN2X scale factors are applied in this validation.

For both VRs, the procedure for the background estimate is retrained and validated. The observed data in the VRs are found to be well described within the assigned background uncertainties, indicating that the background model is sufficiently flexible to account for any residual discrepencies to the real data, as visualized in \cref{fig:postfit-resolved-vr-mS100} and \cref{fig:postfit-mixed-vr-mS750} for two representative resolved and mixed signal masspairs, respectively.
Additionally, signal injection tests were performed injecting the nominal signal in both the Asimov and these `unblinded' VRs. The injected signal was retained by the fit.

\begin{figure}[h!tb]
\centering
\includegraphics[width=\textwidth]{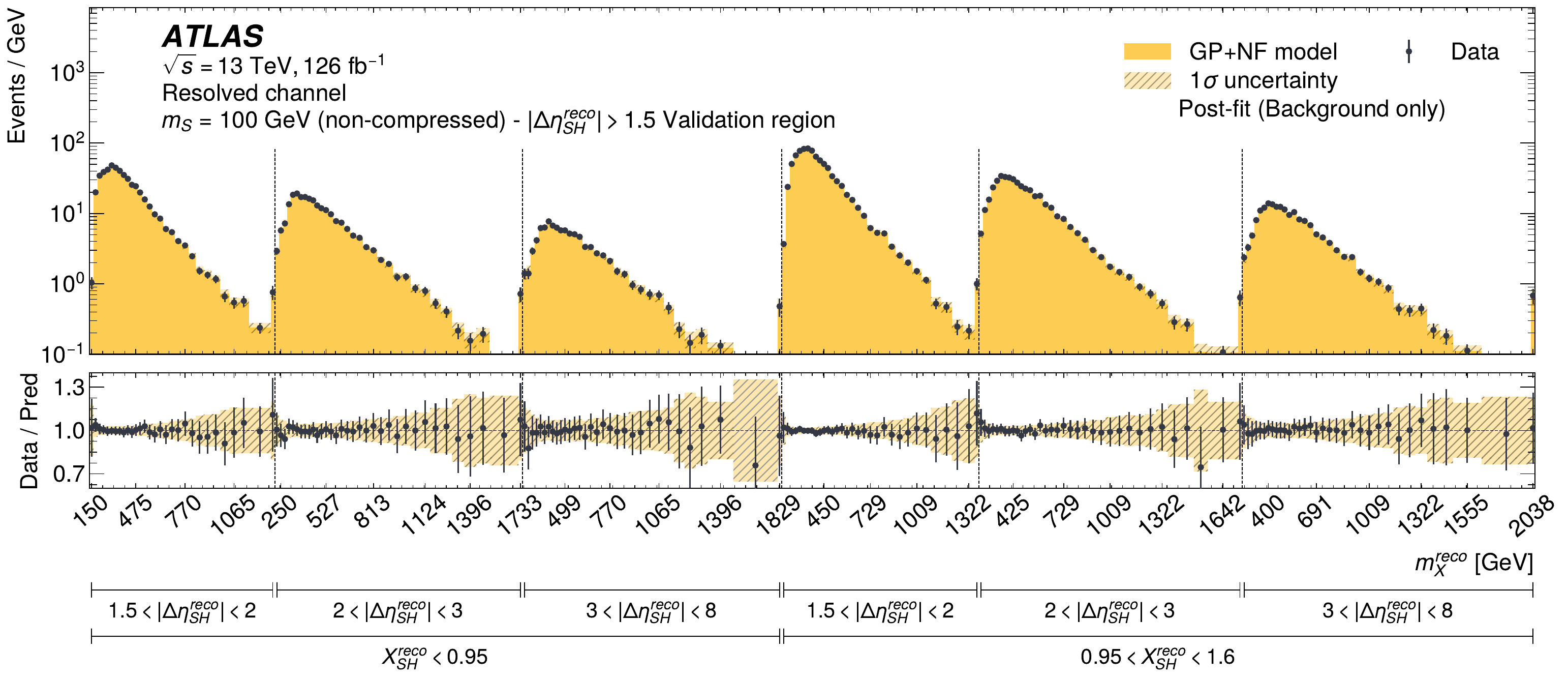}
\caption{Post-fit $m_{\mathrm{X}}^\mathrm{reco}$ distribution in the resolved non-compressed $m_{\mathrm{S}}=100$~\GeV $\Delta \eta_\mathrm{SH}^\mathrm{reco}> 1.5$ validation region for the background-only hypothesis. The lower panel shows the ratio of the observed data to the background prediction. The binning is non-uniform, as the $m_\mathrm{X}^\mathrm{reco}$ distribution is rebinned to ensure a minimum of 10 expected background events per bin.
}
\label{fig:postfit-resolved-vr-mS100}
\end{figure}

\begin{figure}[h!tb]
\centering
\includegraphics[width=0.6\textwidth]{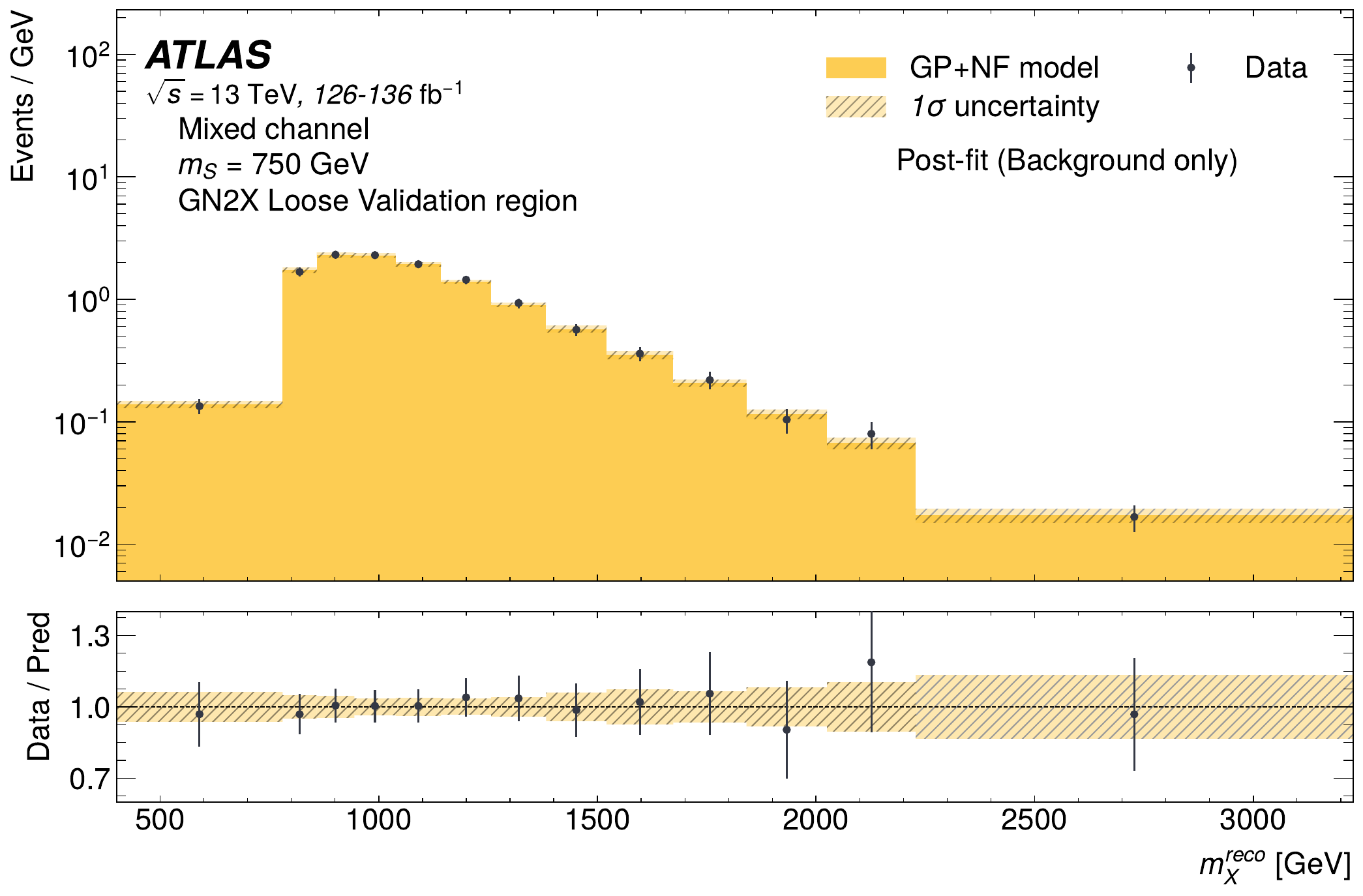}
\caption{Post-fit $m_{\mathrm{X}}^\mathrm{reco}$ distribution in the mixed $m_{\mathrm{S}}=750$~\GeV Xbb-loose validation region for the background-only hypothesis. The lower panel shows the ratio of the observed data to the background prediction. The binning is non-uniform, as the $m_\mathrm{X}^\mathrm{reco}$ distribution is rebinned to ensure a minimum of 10 expected background events per bin.
}
\label{fig:postfit-mixed-vr-mS750}
\end{figure}

% End of text imported from the .//sections/systematics.tex input file

% The next lines are included from the .//sections/results.tex input file
\clearpage
\section{Statistical analysis and results}
\label{sec:results}

\subsection{Statistical analysis}
Upper limits at 95\% CL on the cross-section for the $X \to SH \to 4b$ process are extracted independently for each signal mass pair in the resolved and mixed channels using binned maximum-likelihood fits to the reconstructed $m_{\mathrm{X}}^\mathrm{reco}$ distribution.

The likelihood function is defined by:

\begin{align}
\mathcal{L} \left( \mu, \mathbf{\theta} \right) = \prod_{j=1}^{N} \frac{\left(\mu s_{j} + b_{j} \right)^{n_{j}}}{n_{j} !} e^{-\left(\mu s_{j} + b_{j} \right)} \prod_{k=1}^{M} \frac{ u_{k}^{m_{k}} }{m_{k} !} e^{- u_{k}} .
\label{eq:likelihood}
\end{align}

The first Poisson term describes the probability of observing $n_j$ events in bin $j$, given the expected signal and background yields, where the signal contribution $s_j$ is scaled by the signal-strength parameter $\mu$. The second term encodes constraints on the nuisance parameters $\mathbf{\theta}$ through auxiliary measurements, where $m_k$ denotes the observed auxiliary data and $u_k(\mathbf{\theta})$ its expected value~\cite{Cowan:2010js}.

The statistical model is built from template distributions for the signal and background processes in the signal regions, with $m_{\mathrm{X}}^\mathrm{reco}$ used as the discriminating variable. In the mixed channel, a single signal-region template enters the fit for each signal hypothesis. In the resolved channel, the fit is performed simultaneously across the six categories defined in \cref{sec:selection}, each contributing an independent $m_{\mathrm{X}}^\mathrm{reco}$ template. Systematic uncertainties affecting the signal and background predictions are incorporated as nuisance parameters. Shape uncertainties are implemented through bin-by-bin variations of the templates, while normalization uncertainties are included as multiplicative factors.

The likelihood is maximized with respect to the signal strength parameter $\mu$ and the nuisance parameters $\mathbf{\theta}$, which account for systematic uncertainties. Upper limits on $\mu$ are derived using the profile-likelihood-ratio test statistic, with asymptotic formulae used to evaluate the corresponding confidence intervals~\cite{Cowan:2010js,Read:2002hq}.

The $m_{\mathrm{X}}^\mathrm{reco}$ distribution is rebinned independently for each $m_{\mathrm{S}}$ signal-mass hypothesis to preserve statistical significance across the full mass range and ensure the validity of the asymptotic approximation, requiring a minimum of 10 expected background events per bin.

Studies of the reconstructed mass resolution show that the resolution is significantly better for $m_{\mathrm{S}}^\mathrm{reco}$ than for $m_{\mathrm{X}}^\mathrm{reco}$. The spacing of the tested $m_{\mathrm{S}}$ hypotheses is sufficiently fine that a signal located between neighboring points would produce a measurable excess at one of the tested points for a fixed $m_{\mathrm{X}}$, with at most a moderate loss in sensitivity for the widest intervals, except across the gap between $m_{\mathrm{S}}=100$~\GeV and $m_{\mathrm{S}}=170$~\GeV introduced to avoid overlap with symmetric $HH$ resonances. In contrast, the reconstructed $m_{\mathrm{X}}^\mathrm{reco}$ peaks are substantially narrower, and the current spacing of the $m_{\mathrm{X}}$ hypotheses is not dense enough to support reliable interpolation between neighboring mass points.

\subsection{Results}

The post-fit $m_{\mathrm{X}}^\mathrm{reco}$ distributions for representative signal regions in the resolved and mixed channels are shown in \cref{fig:postfit-resolved-mS100} and \cref{fig:postfit-mixed-mS750}, respectively. The lower panels compare the observed data with the background prediction, and no significant deviation from the background-only expectation is observed.

\begin{figure}[t]
\centering
\includegraphics[width=\textwidth]{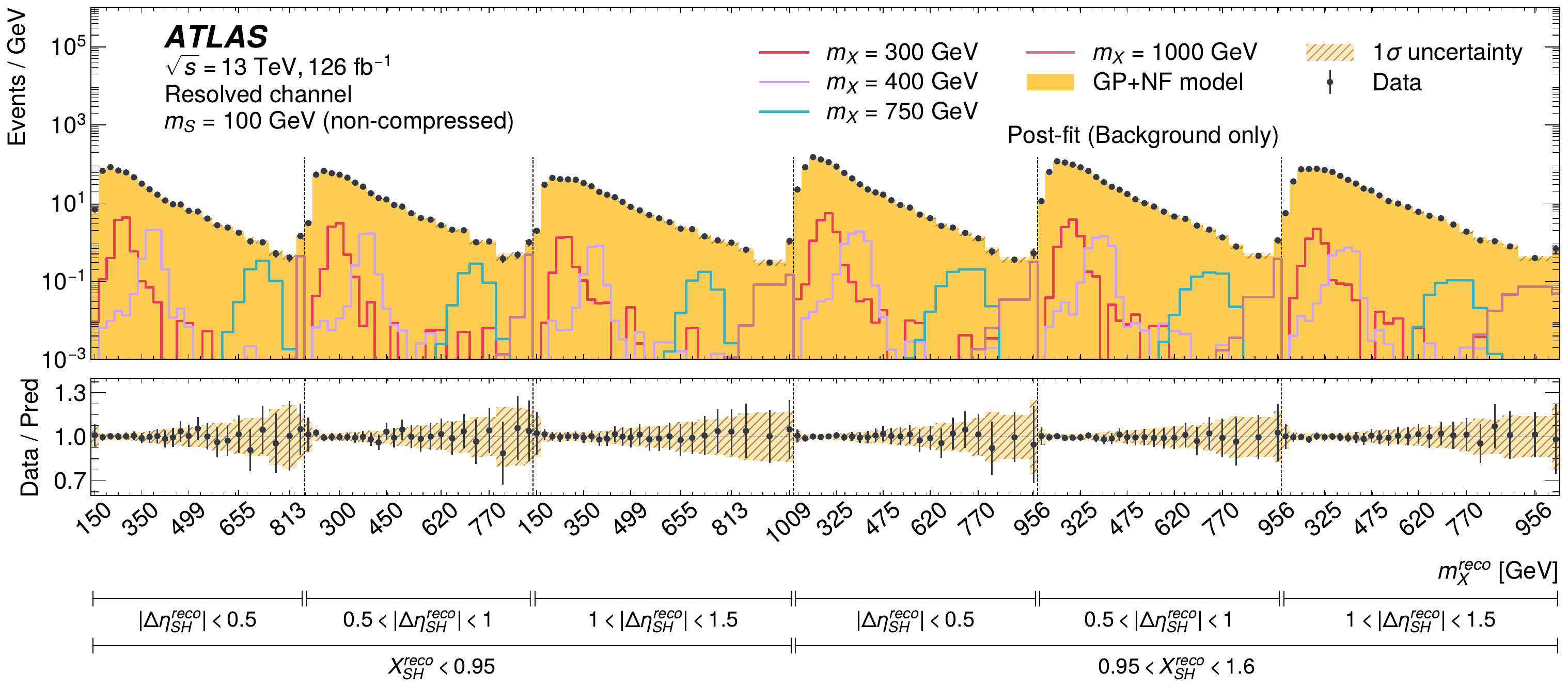}
\caption{Post-fit $m_{\mathrm{X}}^\mathrm{reco}$ distribution in the resolved non-compressed $m_{\mathrm{S}}=100$~\GeV signal region under the background-only hypothesis. Example signal templates for $m_{\mathrm{X}}=400$, 750, and 1000~\GeV are overlaid and scaled to the corresponding expected upper cross-section limit. The lower panel shows the ratio of the observed data to the background prediction.
}
\label{fig:postfit-resolved-mS100}
\end{figure}

\begin{figure}[t]
\centering
\includegraphics[width=0.6\textwidth]{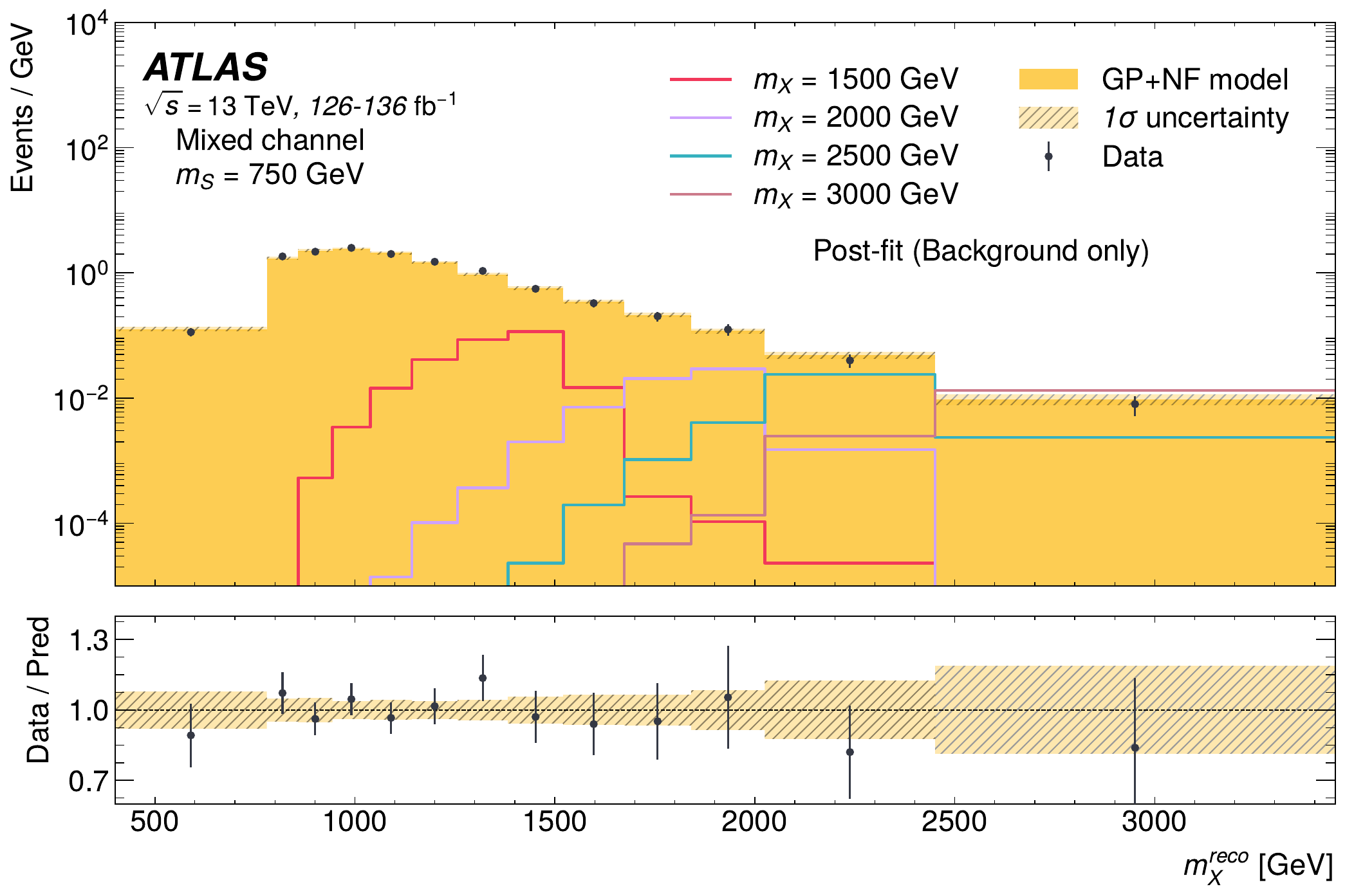}
\caption{Post-fit $m_{\mathrm{X}}^\mathrm{reco}$ distribution in the mixed $m_{\mathrm{S}}=750$~\GeV signal region under the background-only hypothesis. Example signal templates for $m_{\mathrm{X}}=1500$, 2000, 2500, and 3000~\GeV are overlaid and scaled to the corresponding expected upper cross-section limit. The lower panel shows the ratio of the observed data to the background prediction.
}
\label{fig:postfit-mixed-mS750}
\end{figure}

Upper limits at 95\% CL on the signal cross-section times the branching ratio, derived across the full grid of tested mass hypotheses, are shown in~\cref{fig:results_bands}. For mass points covered by both channels, limits are evaluated independently in the resolved and mixed analyses, and the channel providing the stronger expected sensitivity is retained for the final result. These results present a dedicated optimization for $X\to SH\to4b$ resonances with the latest $b$- and $bb$-tagging algorithms, and therefore complement the existing ATLAS resonant searches for symmetric decays $X \to HH\to4b$ \parencite{HDBS-2018-41,HDBS-2022-02}. The obtained upper cross-section limits are being compared with the maximally allowed TRSM prediction in Figure~\ref{fig:results_bands}, approaching sensitivity to this important benchmark for the first time in ATLAS.

\begin{figure}
\center
\includegraphics[width=.9\textwidth]{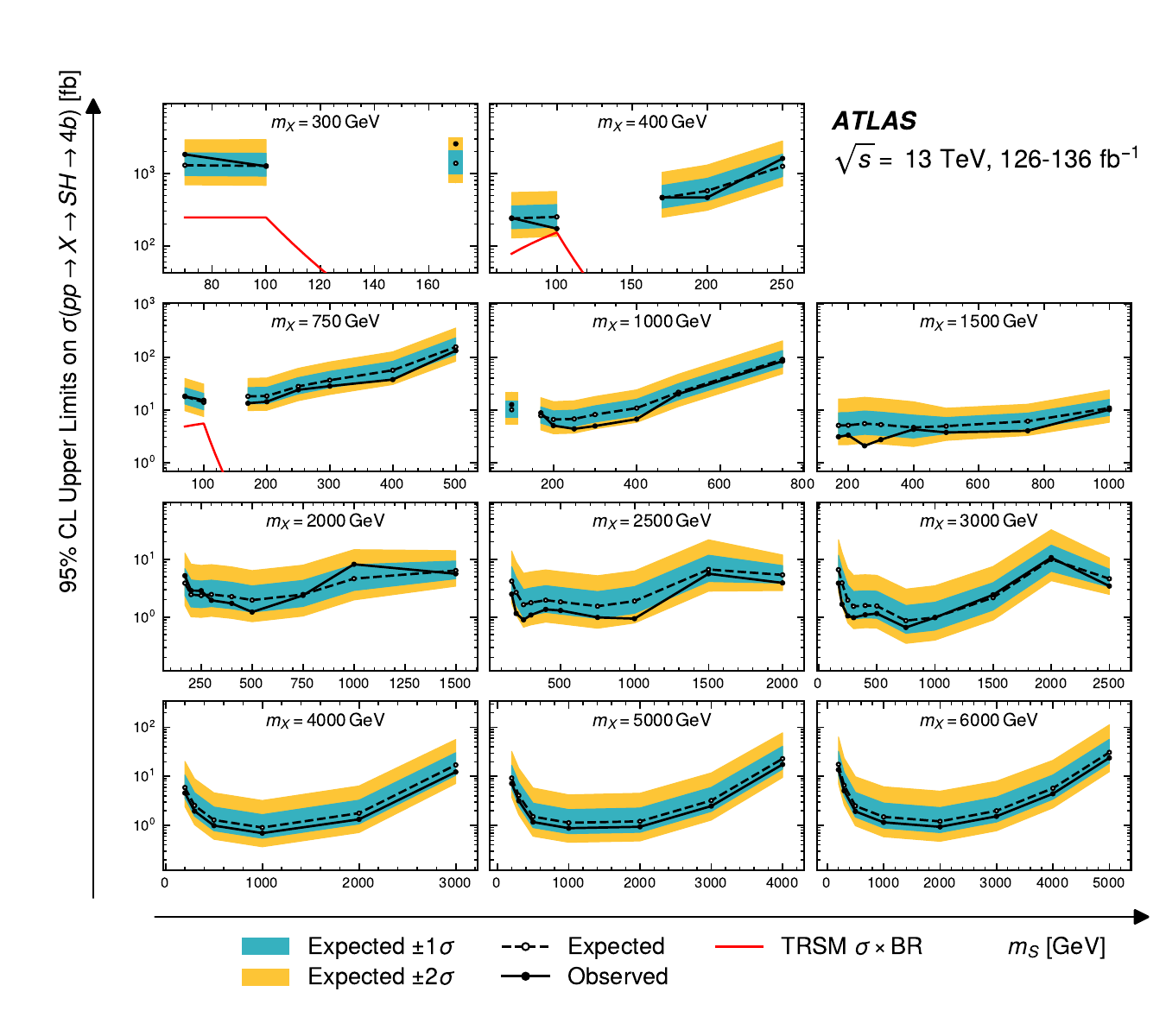}
\caption{Observed and expected 95\% CL upper limits on the $pp \rightarrow X \rightarrow SH \rightarrow 4b$ production cross-section as a function of $m_{\mathrm{S}}$ for each tested $m_{\mathrm{X}}$ hypothesis. For mass hypotheses covered by both the resolved and mixed analyses, only the result from the channel with the lower expected upper limit is retained. The black dashed and solid lines show the median expected and observed upper limits, respectively. The inner teal and outer yellow bands correspond to the $\pm1\sigma$ and $\pm2\sigma$ intervals expected in the absence of signal. The solid red curves show the maximum $\sigma \times \mathcal{B}\left(X \to SH \to 4b\right)$ values allowed in the TRSM parameter space for the corresponding mass hypotheses. The TRSM prediction is shown only where it approaches the experimental sensitivity; for the remaining $m_{\mathrm{X}}$ hypotheses, it lies below the displayed range and is therefore not visible. The red lines between points are illustrative only and do not correspond to theoretical predictions.}
\label{fig:results_bands}
\end{figure}

The strongest observed limits are obtained for the highest tested $m_{\mathrm{X}}$ values, reaching the few-fb level, while weaker limits are found at lower masses where the multijet background is larger. Overall, the observed limits are consistent with the expected sensitivity within uncertainties. The largest excess, with a local significance of $1.8$ standard deviations, is observed at the mass hypothesis $m_{\mathrm{X}}=300~\GeV$, $m_{\mathrm{S}}=170~\GeV$.

Three additional signal mass hypotheses were tested to probe the excesses reported by the CMS Collaboration in their resolved $X\to SH\to b\bar{b}b\bar{b}$ search~\cite{CMS-HIG-20-012}. The results are summarized in Table~\ref{tab:cms_comparison}. At $(m_{\mathrm{X}}, m_{\mathrm{S}}) = (700~\GeV, 400~\GeV)$ and $(1200~\GeV, 900~\GeV)$, the ATLAS observed limits are slightly stronger than those reported by CMS\@. At $(m_{\mathrm{X}}, m_{\mathrm{S}}) = (650~\GeV, 350~\GeV)$, which is the closest ATLAS mass point to the CMS excess at $(600~\GeV, 400~\GeV)$, the sensitivity to the CMS excess is reduced to approximately 70\% relative to a dedicated signal hypothesis due to the shift between the signal $m_{\mathrm{X}}$ value and the hypothesized mass. The ATLAS observed limit is considerably stronger and therefore disfavors the CMS excess.

\begin{table}[htbp]
\centering
\caption{Observed and expected 95\% CL upper limits on  $\sigma \times \mathcal{B}\left(X\to SH \to 4b\right)$ at the three additional mass points tested to probe the excesses reported by the CMS Collaboration.}
\label{tab:cms_comparison}
\begin{tabular}{cccccc}
\toprule
$m_{X}$ [GeV] & $m_{S}$ [GeV] & Observed [fb] & Expected [fb] & Exp.\ $\pm1\sigma$ [fb] & Exp.\ $\pm2\sigma$ [fb] \\
\midrule
650  & 350 & 96.7 & 122.0 & (87.9, 181.4) & (65.5, 278.0) \\
700  & 400 & 62.1 &  96.5 & (69.6, 142.7) & (51.8, 216.5) \\
1200 & 900 & 28.7 &  47.0 & (33.8,\ \ 69.2) & (25.2, 104.6) \\
\bottomrule
\end{tabular}
\end{table}

% End of text imported from the .//sections/results.tex input file

% The next lines are included from the .//sections/conclusions.tex input file
\FloatBarrier
\section{Conclusion}
\label{sec:conclusion}

A search for a heavy scalar resonance $X$ decaying into a new scalar $S$ and a Standard Model-like Higgs boson $H$ is presented in the $4b$ final state. The analysis uses the full Run 2 proton--proton collision data at $\sqrt{s}=13$~\TeV recorded by the ATLAS detector at the LHC, corresponding to an integrated luminosity of $126$--$136~\ifb$. The search probes a broad region of the $(m_{\mathrm{X}},m_{\mathrm{S}})$ plane, with masses in the ranges $300~\GeV \leq m_{\mathrm{X}} \leq 6~\TeV$ and $70~\GeV \leq m_{\mathrm{S}} \leq 5~\TeV$. Two complementary event topologies are considered: a resolved channel, in which both bosons are reconstructed from pairs of small-radius jets, and a mixed channel, in which the Higgs boson is reconstructed as a boosted large-radius jet while the $S$ boson is reconstructed from two small-radius jets. Novel data-driven techniques for the background estimate are employed, based on Gaussian processes and normalizing flows.

Upper limits at 95\% confidence level are set on the production cross-section multiplied by the branching ratio for the $pp \rightarrow X \rightarrow SH \rightarrow 4b$ process. Depending on the tested mass hypothesis, the observed limits range from $0.7$~fb to $2.6$~pb. As the multijet background decreases rapidly with increasing resonance mass, the strongest limits are obtained at the highest values of $m_{\mathrm{X}}$. No significant deviation from the Standard Model expectation is observed. The largest excess, with a local significance of $1.8~\sigma$, is observed at the mass hypothesis $m_{\mathrm{X}}=300$~\GeV, $m_{\mathrm{S}}=170$~\GeV. These results provide the first ATLAS constraints on the $X \rightarrow SH \rightarrow 4b$ signature and significantly extend the experimental sensitivity to asymmetric scalar resonance decays in fully hadronic final states.

% End of text imported from the .//sections/conclusions.tex input file

\FloatBarrier

\section*{Acknowledgements}
%
%

% The next lines are included from the .//acknowledgements/Acknowledgements.tex input file
%
%

%
%

We thank CERN for the very successful operation of the LHC and its injectors, as well as the support staff at
CERN and at our institutions worldwide without whom ATLAS could not be operated efficiently.

The crucial computing support from all WLCG partners is acknowledged gratefully, in particular from CERN, the ATLAS Tier-1 facilities at TRIUMF/SFU (Canada), NDGF (Denmark, Norway, Sweden), CC-IN2P3 (France), KIT/GridKA (Germany), INFN-CNAF (Italy), NL-T1 (Netherlands), PIC (Spain), RAL (UK) and BNL (USA), the Tier-2 facilities worldwide and large non-WLCG resource providers. Major contributors of computing resources are listed in Ref.~\cite{ATL-SOFT-PUB-2026-001}.

We gratefully acknowledge the support of ANPCyT, Argentina; YerPhI, Armenia; ARC, Australia; BMWFW and FWF, Austria; ANAS, Azerbaijan; CNPq and FAPESP, Brazil; NSERC, NRC and CFI, Canada; CERN; ANID, Chile; CAS, MOST, NSFC and FRFCU, China; Minciencias, Colombia; MEYS CR, Czech Republic; DNRF and DNSRC, Denmark; IN2P3-CNRS and CEA-DRF/IRFU, France; SRNSFG, Georgia; BMFTR, HGF and MPG, Germany; GSRI, Greece; RGC and Hong Kong SAR, China; ICHEP and Academy of Sciences and Humanities, Israel; INFN, Italy; MEXT and JSPS, Japan; CNRST, Morocco; NWO, Netherlands; RCN, Norway; MNiSW, Poland; FCT, Portugal; MNE/IFA, Romania; MSTDI, Serbia; MSSR, Slovakia; ARIS and MVZI, Slovenia; DSI/NRF, South Africa; MICIU/AEI, Spain; SRC and Wallenberg Foundation, Sweden; SERI, SNSF and Cantons of Bern and Geneva, Switzerland; NSTC, Taipei; TENMAK, T\"urkiye; STFC/UKRI, United Kingdom; DOE and NSF, United States of America.

Individual groups and members have received support from BCKDF, CANARIE, CRC and DRAC, Canada; CERN-CZ, FORTE and PRIMUS, Czech Republic; COST, ERC, ERDF, Horizon 2020 and Marie Sk{\l}odowska-Curie Actions, European Union; Investissements d'Avenir Labex, Investissements d'Avenir Idex and ANR, France; DFG and AvH Foundation, Germany; Herakleitos, Thales and Aristeia programmes co-financed by EU-ESF and the Greek NSRF, Greece; BSF-NSF and MINERVA, Israel; NCN and NAWA, Poland; La Caixa Banking Foundation, CERCA and AGAUR programs from Generalitat de Catalunya and PROMETEO and GenT Programmes Generalitat Valenciana, Spain; G\"{o}ran Gustafssons Stiftelse, Sweden; The Royal Society and Leverhulme Trust, United Kingdom; Eric and Wendy Schmidt Fund for Strategic Innovation, United States of America.

In addition, individual members wish to acknowledge support from Chile: Agencia Nacional de Investigaci\'on y Desarrollo (ANID FONDECYT reg. 1230987, FONDECYT 1230812, FONDECYT 1240864, Fondecyt 3240661, Fondecyt Regular 1240721); China: Fundamental Research Funds for the Central Universities (010-63263105), Chinese Ministry of Science and Technology (MOST-2023YFA1605700, MOST-2023YFA1609300), National Natural Science Foundation of China (NSFC 12275265, NSFC-W2543005); Czech Republic: Czech Science Foundation (GACR - 24-11373S), Ministry of Education Youth and Sports (ERC-CZ-LL2327, FORTE CZ.02.01.01/00/22\_008/0004632); EU: H2020 European Research Council (ERC - 101002463); European Union: European Research Council (BARD No. 101116429, ERC - 101219398, ERC 101089007), European Regional Development Fund (HE COFUND GA No.101081355, ERDF), Marie Sklodowska-Curie Actions (GAP-101168829); France: Agence Nationale de la Recherche (ANR-22-EDIR-0002, ANR-24-CE31-0504-01); Germany: Deutsche Forschungsgemeinschaft (DFG - 469666862); China: Research Grants Council (GRF); Italy: Istituto Nazionale di Fisica Nucleare (LHC-MIUR - 28003/2025), Ministero dell'Università e della Ricerca (NextGenEU I53D23000820006 M4C2.1.1, SOE2024\_0000023); Japan: Japan Society for the Promotion of Science (JSPS KAKENHI  JP25H0063, JSPS KAKENHI JP22H04944, JSPS KAKENHI JP24K23939, JSPS KAKENHI JP24KK0251, JSPS KAKENHI JP25H00650, JSPS KAKENHI JP25H01291, JSPS KAKENHI JP25K01011, JSPS KAKENHI JP25K01023, JSPS KAKENHI JP25KK0047); Poland: Polish National Science Centre (NCN 2021/42/E/ST2/00350, NCN OPUS 2023/51/B/ST2/02507, NCN OPUS nr 2022/47/B/ST2/03059, NCN UMO-2019/34/E/ST2/00393, UMO-2022/47/O/ST2/00148, UMO-2023/49/B/ST2/04085, UMO-2023/51/B/ST2/00920, UMO-2024/53/N/ST2/00869); Spain: Agència de Gestió d'Ajuts Universitaris i de Recerca. (AGAUR - 2023 BP 00141), Ministry of Science and Innovation (RYC2019-028510-I, RYC2020-030254-I, RYC2021-031273-I, RYC2022-038164-I), Ministerio de Ciencia, Innovación y Universidades/Agencia Estatal de Investigaci\'on (EU NextGenerationEU (PRTR-C17.I1), PID2022-142604OB-C22); Sweden: Carl Trygger Foundation (Carl Trygger Foundation CTS 22:2312), Swedish Research Council (Swedish Research Council 2023-04654, VR 2021-03651, VR 2022-03845, VR 2022-04683, VR 2023-03403, VR 2024-05451, VR 2025-05940), Knut and Alice Wallenberg Foundation (KAW 2023.0366); United Kingdom: The Binks Trust, Royal Society (NIF-R1-231091); United States of America: U.S. Department of Energy (ECA DE-AC02-76SF00515), John Templeton Foundation (John Templeton Foundation 63206), Neubauer Family Foundation.

%
%

% End of text imported from the .//acknowledgements/Acknowledgements.tex input file

%

\clearpage
\appendix
\part*{Appendix}
\addcontentsline{toc}{part}{Appendix}

% The next lines are included from the .//sections/app_trsm.tex input file
\section{TRSM predictions}
\label{app:trsm}

The maximally allowed gluon--gluon fusion cross-sections in the TRSM~\cite{Robens_2020} for $\sqrt{s}=13~\TeV$, using the latest theoretical and experimental bounds~\cite{Bahl_2023} and considering the experimental mass resolution, are summarized in Table~\ref{tab:trsm_values}, along with the values of the model parameters~\cite{trsmscans}.

\begin{table}[h!]
\tiny
\centering
\caption{Maximized ggF cross-section times branching ratio, $\sigma\times \mathcal{B}\left(X\to SH\to 4b\right)$, and the corresponding theory parameter values, for $\sqrt{s}=13~\TeV$. The mixing angles are $\theta_{HS}$, $\theta_{HX}$ and $\theta_{SX}$. The vacuum expectation values are called $v_S$ and $v_S$. The branching ratios for $X\to SH$ and $S\to b\bar{b}$ are stated for information. The predicted widths of the $X$ and the $S$ are labeled $\Gamma_S$ and $\Gamma_X$. The width of the $H$ is almost constant with a value of approximately 3.8~\MeV. The branching ratio of $H\to b\bar{b}$ has a constant value of 59\%.}
\label{tab:trsm_values}

\begin{tabular}{c c c c c c c c c c c c}
\toprule
$m_X$ & $m_S$ & $\sigma\times \mathcal{B}$ & $\mathcal{B}\left(X\to SH\right)$ & $\mathcal{B}\left(S\to b\bar{b}\right)$ & $\theta_{HS}$ & $\theta_{HX}$ & $\theta_{SX}$ & $v_X$ & $v_S$ & $\Gamma_X/m_X$ & $\Gamma_S/m_S$ \\
{}[GeV] & [GeV] & [fb] & [\%] & [\%] &  &  &  & [GeV] & [GeV] & [\%] & [\%] \\
\midrule

300 & 70 & 247 & 73.4 & 84.2 & 1.49 & 0.681 & -0.318 & 78.9 & 129 & 1.03 & <0.001 \\
300 & 100 & 246 & 59.9 & 79.8 & 1.32 & 1.1 & 0.0819 & 351 & 81.6 & 0.832 & <0.001 \\
300 & 170 & 2.20 & 62.5 & 0.858 & -0.26 & 0.112 & -0.868 & 279 & 63.0 & 0.537 & 0.001 \\
400 & 70 & 77.7 & 42.3 & 84.2 & 1.39 & 1.05 & -0.357 & 151 & 125 & 0.786 & <0.001 \\
400 & 100 & 153 & 49.7 & 79.8 & -1.39 & -0.783 & 0.14 & 124 & 94.8 & 1.82 & <0.001 \\
400 & 170 & 2.19 & 60.4 & 0.858 & -0.147 & 0.243 & -0.757 & 89.7 & 91.0 & 2.23 & 0.001 \\
400 & 200 & 0.646 & 69.7 & 0.257 & -0.256 & 0.108 & -0.795 & 176 & 81.2 & 1.63 & 0.007 \\
400 & 250 & 0.198 & 59.4 & 0.107 & 0.143 & -0.199 & -0.863 & 86.3 & 111 & 1.21 & 0.005 \\
750 & 70 & 4.86 & 21.5 & 84.2 & -1.4 & 0.455 & 0.35 & 157 & 890 & 4.28 & <0.001 \\
750 & 100 & 5.55 & 18.9 & 79.8 & 1.28 & -1.04 & -0.064 & 805 & 172 & 5.41 & <0.001 \\
750 & 170 & 0.0793 & 29.4 & 0.858 & 0.202 & -0.249 & -1.1 & 187 & 184 & 4.77 & 0.003 \\
750 & 200 & 0.0258 & 30.4 & 0.257 & -0.225 & 0.209 & -0.646 & 172 & 200 & 4.84 & 0.002 \\
750 & 250 & 0.0151 & 40.9 & 0.107 & 0.191 & -0.246 & -0.811 & 150 & 166 & 5.51 & 0.003 \\
750 & 300 & 0.00743 & 44.9 & 0.058 & 0.147 & -0.254 & -0.833 & 141 & 175 & 4.93 & 0.021 \\
750 & 400 & 0.00277 & 43.1 & 0.022 & -0.255 & 0.159 & -0.705 & 228 & 136 & 4.78 & 0.062 \\
750 & 500 & 0.00125 & 36.5 & 0.012 & 0.14 & -0.265 & -0.812 & 160 & 231 & 4.20 & 0.110 \\
1000 & 70 & 0.569 & 17.9 & 84.2 & -1.38 & 0.507 & 0.33 & 207 & 870 & 5.80 & <0.001 \\
1000 & 100 & 0.561 & 21.0 & 79.8 & 1.09 & -1.04 & 0.196 & 677 & 241 & 5.10 & <0.001 \\
1000 & 170 & 0.00625 & 16.4 & 0.858 & -0.219 & 0.192 & -0.337 & 257 & 154 & 6.70 & 0.004 \\
1000 & 200 & 0.00215 & 16.1 & 0.257 & -0.231 & 0.202 & -0.39 & 277 & 210 & 6.57 & 0.013 \\
1000 & 250 & 0.00107 & 20.0 & 0.107 & -0.216 & 0.182 & -0.544 & 241 & 258 & 5.93 & 0.013 \\
1000 & 300 & 0.00067 & 25.3 & 0.056 & 0.156 & -0.23 & -0.974 & 238 & 236 & 5.75 & 0.028 \\
1000 & 400 & 0.000307 & 27.9 & 0.022 & 0.139 & -0.24 & -0.799 & 220 & 240 & 5.96 & 0.037 \\
1000 & 500 & 0.000183 & 33.5 & 0.011 & 0.122 & -0.255 & -0.836 & 193 & 257 & 6.06 & 0.142 \\
1000 & 750 & 6.82e-05 & 23.1 & 0.005 & 0.163 & -0.265 & -0.855 & 221 & 284 & 6.82 & 0.232 \\
1500 & 70 & 0.00913 & 9.02 & 84.2 & 1.32 & -0.909 & 0.0459 & 905 & 372 & 6.88 & <0.001 \\
1500 & 100 & 0.0148 & 10.1 & 79.8 & 0.513 & 1.29 & -1.23 & 133 & 478 & 6.67 & <0.001 \\
1500 & 170 & 0.000192 & 16.6 & 0.858 & 0.157 & -0.251 & 0.418 & 366 & 849 & 5.73 & 0.012 \\
1500 & 200 & 3.93e-05 & 11.3 & 0.257 & 0.129 & -0.23 & 0.39 & 431 & 906 & 5.14 & 0.029 \\
1500 & 250 & 1.53e-05 & 10.2 & 0.107 & -0.199 & 0.116 & -0.279 & 412 & 256 & 5.78 & 0.040 \\
1500 & 300 & 8.33e-06 & 13.3 & 0.046 & -0.213 & 0.095 & -0.356 & 387 & 293 & 5.59 & 0.097 \\
1500 & 400 & 3.62e-06 & 14.2 & 0.018 & -0.202 & 0.0877 & -0.46 & 350 & 360 & 5.34 & 0.176 \\
1500 & 500 & 2.13e-06 & 12.2 & 0.012 & -0.132 & 0.11 & -0.777 & 380 & 314 & 5.19 & 0.004 \\
1500 & 750 & 1.94e-06 & 32.2 & 0.004 & 0.088 & -0.268 & 0.587 & 326 & 364 & 5.89 & 1.71 \\
1500 & 1000 & 1.02e-06 & 32.8 & 0.002 & 0.0745 & -0.273 & 0.669 & 325 & 442 & 5.53 & 3.11 \\
2000 & 70 & 0.000539 & 3.85 & 84.2 & 0.543 & 1.39 & -1.18 & 114 & 908 & 9.39 & <0.001 \\
2000 & 100 & 0.00119 & 19.3 & 79.8 & -1.2 & 0.45 & 0.285 & 852 & 828 & 5.58 & <0.001 \\
2000 & 170 & 1.08e-05 & 17.1 & 0.858 & 0.211 & -0.219 & 0.424 & 583 & 774 & 5.81 & 0.016 \\
2000 & 200 & 2.52e-06 & 9.81 & 0.257 & 0.158 & -0.186 & 0.315 & 498 & 691 & 6.33 & 0.029 \\
2000 & 250 & 8.78e-07 & 8.18 & 0.107 & 0.126 & -0.179 & 0.337 & 481 & 741 & 6.43 & 0.049 \\
2000 & 300 & 3.03e-07 & 9.69 & 0.045 & 0.0623 & -0.207 & -1.36 & 292 & 552 & 4.90 & 0.127 \\
2000 & 400 & 9.89e-08 & 7.61 & 0.017 & -0.172 & 0.0642 & -0.289 & 509 & 388 & 4.84 & 0.195 \\
2000 & 500 & 6.24e-08 & 8.14 & 0.009 & 0.0576 & -0.176 & -1.22 & 461 & 543 & 4.98 & 0.335 \\
2000 & 750 & 5.33e-08 & 16.2 & 0.004 & 0.0375 & -0.274 & -1.28 & 353 & 646 & 5.58 & 2.13 \\
2000 & 1000 & 6.6e-08 & 42.2 & 0.002 & 0.17 & -0.237 & 0.549 & 521 & 378 & 5.48 & 4.60 \\
2000 & 1500 & 2.78e-08 & 35.8 & 0.001 & 0.123 & -0.243 & 0.68 & 490 & 445 & 4.90 & 9.01 \\
2500 & 70 & 4.41e-05 & 6.44 & 84.183 & -1.38 & 0.451 & 0.172 & 797 & 876 & 6.61 & <0.001 \\
2500 & 100 & 0.000133 & 20.1 & 79.826 & -1.22 & 0.471 & 0.255 & 570 & 922 & 7.94 & <0.001 \\
2500 & 170 & 9.41e-07 & 13.0 & 0.858 & -0.165 & 0.236 & 1.27 & 599 & 723 & 6.03 & 0.015 \\
2500 & 200 & 2.19e-07 & 8.71 & 0.257 & 0.158 & -0.16 & 0.335 & 604 & 814 & 6.84 & 0.028 \\
2500 & 250 & 5.43e-08 & 6.99 & 0.107 & 0.145 & -0.136 & 0.312 & 643 & 947 & 4.93 & 0.050 \\
2500 & 300 & 2.07e-08 & 9.33 & 0.044 & 0.0305 & -0.204 & -1.38 & 303 & 755 & 4.51 & 0.136 \\
2500 & 400 & 5.55e-09 & 5.34 & 0.015 & -0.116 & 0.136 & 1.33 & 562 & 695 & 4.83 & 0.254 \\
2500 & 500 & 4.11e-09 & 5.74 & 0.009 & 0.0432 & -0.162 & -1.29 & 497 & 718 & 5.13 & 0.330 \\
2500 & 750 & 3.84e-09 & 12.2 & 0.004 & -0.277 & 0.0201 & -0.253 & 705 & 521 & 4.91 & 2.39 \\
2500 & 1000 & 2.24e-09 & 12.2 & 0.002 & 0.0305 & -0.269 & -1.34 & 463 & 753 & 5.30 & 3.88 \\
2500 & 1500 & 2.89e-09 & 39.5 & 0.001 & -0.236 & 0.136 & 0.825 & 713 & 544 & 5.20 & 9.80 \\
2500 & 2000 & 1.43e-09 & 21.8 & 0.001 & 0.107 & -0.224 & 0.702 & 614 & 630 & 6.13 & 14.6 \\

\bottomrule
\end{tabular}
\end{table}

% End of text imported from the .//sections/app_trsm.tex input file

%
%
%

%
%
%
\printbibliography
\clearpage
% ATLAS Collaboration author list
% Reference date of HIGP-2024-31 is 2026-06-12
% Author list last updated on date 20-JUL-26
% Data extracted on 20-Jul-2026 for paper reference HIGP-2024-31
% at 9:00pm
 
\begin{flushleft}
\hypersetup{urlcolor=black}
{\Large The ATLAS Collaboration}

\bigskip

\AtlasOrcid[0000-0002-6665-4934]{G.~Aad}$^\textrm{\scriptsize 102}$,
\AtlasOrcid[0000-0001-7616-1554]{E.~Aakvaag}$^\textrm{\scriptsize 17}$,
\AtlasOrcid[0000-0002-5888-2734]{B.~Abbott}$^\textrm{\scriptsize 121}$,
\AtlasOrcid[0000-0002-0287-5869]{S.~Abdelhameed}$^\textrm{\scriptsize 83b}$,
\AtlasOrcid[0000-0002-1002-1652]{K.~Abeling}$^\textrm{\scriptsize 54}$,
\AtlasOrcid[0000-0001-5763-2760]{N.J.~Abicht}$^\textrm{\scriptsize 48}$,
\AtlasOrcid[0000-0002-8496-9294]{S.H.~Abidi}$^\textrm{\scriptsize 30}$,
\AtlasOrcid[0009-0003-6578-220X]{M.~Aboelela}$^\textrm{\scriptsize 44}$,
\AtlasOrcid[0000-0002-9987-2292]{A.~Aboulhorma}$^\textrm{\scriptsize 36e}$,
\AtlasOrcid[0000-0001-5329-6640]{H.~Abramowicz}$^\textrm{\scriptsize 154}$
% ...
\bigskip
\\

% ...
$^{17}$Department for Physics and Technology, University of Bergen, Bergen; Norway.\\
% ...
$^{30}$Physics Department, Brookhaven National Laboratory, Upton NY; United States of America.\\
% ...
$^{36}$$^{(a)}$Facult\'e des Sciences Ain Chock, Universit\'e Hassan II de Casablanca;$^{(b)}$Facult\'{e} des Sciences, Universit\'{e} Ibn-Tofail, K\'{e}nitra;$^{(c)}$Facult\'e des Sciences Semlalia, Universit\'e Cadi Ayyad, LPHEA-Marrakech;$^{(d)}$LPMR, Facult\'e des Sciences, Universit\'e Mohamed Premier, Oujda;$^{(e)}$Facult\'e des sciences, Universit\'e Mohammed V, Rabat;$^{(f)}$Institute of Applied Physics, Mohammed VI Polytechnic University, Ben Guerir; Morocco.\\
% ...
$^{44}$Physics Department, Southern Methodist University, Dallas TX; United States of America.\\
% ...
$^{48}$Fakult\"{a}t Physik, Technische Universit{\"a}t Dortmund, Dortmund; Germany.\\
% ...
$^{54}$II. Physikalisches Institut, Georg-August-Universit\"{a}t G\"ottingen, G\"ottingen; Germany.\\
% ...
$^{83}$$^{(a)}$Khalifa University of Science and Technology, Abu Dhabi;$^{(b)}$New York University Abu Dhabi, Abu Dhabi;$^{(c)}$United Arab Emirates University, Al Ain;$^{(d)}$University of Sharjah, Sharjah; United Arab Emirates.\\
% ...
$^{102}$CPPM, Aix-Marseille Universit\'e, CNRS/IN2P3, Marseille; France.\\
% ...
$^{121}$Homer L. Dodge Department of Physics and Astronomy, University of Oklahoma, Norman OK; United States of America.\\
% ...
$^{154}$Raymond and Beverly Sackler School of Physics and Astronomy, Tel Aviv University, Tel Aviv; Israel.\\
% ...
$^{a}$ Also at Affiliated with an institute formerly covered by a cooperation agreement with CERN.\\
$^{b}$ Also at An-Najah National University, Nablus; Palestine.\\
$^{c}$ Also at Borough of Manhattan Community College, City University of New York, New York NY; United States of America.\\
$^{d}$ Also at Center for Interdisciplinary Research and Innovation (CIRI-AUTH), Thessaloniki; Greece.\\
$^{e}$ Also at Centre of Physics of the Universities of Minho and Porto (CF-UM-UP); Portugal.\\
$^{f}$ Also at CERN, Geneva; Switzerland.\\
% ...

\end{flushleft}

% Created with Glance <Atlas.Glance@cern.ch>

%

%
%

%

%
%
%
%

\end{document}